\documentstyle[12pt,aasms4]{article}
\def\yrm1{yr$^{-1}$}
\def\sun{\odot}
\def\msun{$M_{\sun}$~}

\def\msunend{$M_{\sun}$}

\def\mch{$M_{\rm Ch}$~}

\def\mche{$M_{\rm Ch}$~}

\def\dash{------}

\def\tdfloc{$\tau_{\rm diff,loc}$~}
\def\tchloc{$\tau_{\rm ch,loc}$~}
\def\tevo{$\tau_{\rm evo}$~}
\def\tch{$\tau_{\rm ch}$~}
\def\tchend{$\tau_{\rm ch}$}
\def\tdf{$\tau_{\rm diff}$~}
\def\tdfend{$\tau_{\rm diff}$}
\def\mdot{$\dot{M}$~}
\def\mdotend{$\dot{M}$}
\def\o0{$\omega_{0}$~}
\def\oth{$\omega_{\rm th}$~}
\begin{document}
\title{
Carbon-Oxygen White Dwarfs Accreting CO-Rich Matter I: A Comparison Between 
Rotating and Non-Rotating Models}
\author{Luciano Piersanti  \altaffilmark{1}}
\affil{Dipartimento di Fisica dell'Universit\`a degli Studi di Napoli
``Federico II'', Mostra d'Oltremare, pad. 20, 80125, Napoli, Italy; 
piersanti@astrte.te.astro.it}

\author{Simona Gagliardi}
\affil{Osservatorio Astronomico di Teramo, Via M.Maggini 47, 
64100 Teramo, Italy; gagliardi@astrte.te.astro.it} 

\author{Icko Iben Jr.}
\affil{Astronomy Department, University of Illinois, 
1002 W. Green Street, Urbana, IL 61801;icko@astro.uiuc.edu}

\and

\author{Amedeo Tornamb\'e
\affil{Osservatorio Astronomico di Teramo, Via M.Maggini 47, 
64100 Teramo, Italy; tornambe@astrte.te.astro.it}
}
\noindent
\altaffilmark{1}{Osservatorio Astronomico di Teramo, Via M.Maggini 
47, 64100 Teramo, Italy} 

\begin{abstract}

We investigate the lifting effect of rotation on the thermal evolution
of carbon-oxygen white dwarfs accreting carbon-oxygen-rich matter. We find that
rotation induces the cooling of the accreting structure so that the delivered 
gravitational energy causes a greater expansion with respect to the standard
non-rotating case. The increase in the surface radius produces a decrease in 
the surface value of the critical angular velocity and, therefore, 
the accreting white dwarf becomes gravitationally unbound (Roche instability). 
This occurrence is due to an increase in the total angular momentum of the
accreting white dwarf and depends critically on
the amount of specific angular momentum deposited by the accreted matter. If
the specific angular momentum of the accreted matter is equal to that of 
the outer layers of the accreting structure, the Roche instability occurs 
well before the accreting white dwarf can attain the physical conditions for
carbon burning. If the values of both initial angular 
velocity and accretion rate are small, we find that the accreting white dwarf 
undergoes a secular instability when its total mass approaches 1.4 \msun .
At this stage, the ratio between the rotational energy and  the
gravitational binding energy of the white dwarf
becomes of the order of 0.1, so that the star must deform by adopting an
elliptical shape. In this case, since the angular velocity of the
white dwarf is as large as $\sim 1$ rad s$^{-1}$, the anisotropic mass
distribution induces the loss of rotational energy and angular momentum
via gravitational wave radiation. 
We find that, independent of the braking efficiency, the white dwarf
contracts and achieves the physical conditions suitable for explosive
carbon burning at the center so that a type Ia supernova event is produced.

\end{abstract}

{\em Subject headings}: stars: accretion, rotation - supernovae: general 
- white dwarf

\section{Introduction}

Type Ia Supernovae (SNe Ia) are one of the most powerful tools for
determining the rate of expansion and elucidating the geometrical
structure of the Universe. They are intrinsically
very luminous objects and, provided the empirical relation between the
absolute magnitude at maximum light and the rate of decline from maximum
(Phillips 1993) proves to be theoretically well founded and universally valid,
they may eventually be used as fairly reliable distance indicators. For
each observed event, it may then be possible to estimate reliably the
luminosity at maximum and, hence, to determine the distance modulus. 

Despite the pivotal role of the observed correlation in observational
cosmology, a theoretical justification for the correlation does not yet exist.
Nevertheless, a Phillips relation which has been constructed from a sample of
only nine local ($z < 0.5$) supernovae (SNe) has been used to estimate
the distances to many more SNe at much larger $z$, assuming that no differences
at all exist between events produced by different stellar populations in
the past.

Finally, from the theoretical point of view, a definitive identification
of the progenitor system and an understanding the full nature of the explosion
mechanism are still missing. 
At present it is accepted that SNe Ia are produced by the thermonuclear
disruption of a carbon-oxygen (CO) white dwarf (WD) which has attained
the Chandrasekhar mass (\mche). Hoyle \& Fowler (1960) were the
first to examine the nucleosynthetic characteristics of such an explosion,
without specifying how the initial configuration comes about. Arnett (1969)
demonstrated that, without the mass loss which in real life terminates
its evolution, an intermediate mass single star would ultimately develop
a CO core of Chandrasekhar mass. We now know that single intermediate
mass stars end their nuclear burning evolution with the ejection of their
hydrogen-rich envelopes (illuminated for a brief time as planetary nebulae)
and evolve into white dwarfs of mass considerably
smaller than the Chandrasekhar mass. The consensus view has become that
the progenitors of type Ia supernovae are close binary star systems.
The most traditional scenario is the single degenerate (SD) one in which
it is assumed that the accreted matter is supplied by a companion which
is composed of hydrogen-rich matter (Whelan \& Iben 1973). An alternate
scenario is the double-degenerate (DD)
one which argues that evolutionary paths exist for producing two CO
white dwarfs which are close enough to merge in a Hubble time (due to
angular momentum loss by gravitational wave radiation) and whose combined
mass exceeds \mche. This second scenario, examined by Iben \& Tutukov (1984)
and by Webbink (1984), has been questioned from time to time. 
Two main features make the DD scenario attractive: 
1) the total absence of hydrogen in the progenitor system and 
2) the fact that the more massive WD approaches the critical mass in a very 
natural way. Yet, in spite of huge observational efforts, only one good
candidate SN Ia progenitor has been identified (KPD 0422+5421 - see
Koen, Orosz \& Wade 1998); in fact, this is the only system yet found
the total mass of which is of the order of \mch and the components of which
will merge in less than a Hubble time due to gravitational wave radiation
(GWR). Other observed systems with an orbital period short enough for a
merger to occur in less than a Hubble time have a total mass which is too
small for the system to qualify as a SN Ia progenitor
(see Saffer, Livio \& Yungelson (1998) and also Livio (2000) for a recent
review). Nevertheless, population synthesis estimates suggest
that enough massive close binary systems are produced in nature 
to account for the observed frequency of SNe Ia events
(Iben, Tutukov, and Yungelson (1997); Livio (2000), and references therein).
Most of these systems are undetectable because of their very low intrinsic
luminosity. However, thanks to an increase in luminosity due to viscous heating
(Webbink and Iben 1988; Iben, Tutukov, and Fedorova 1998), perhaps a very
few can be seen just before merger, due to an occultation-modulated light
curve.

It is commonly understood (e.g., Tutukov and Yungelson (1979), Iben and
Tutukov (1984), Benz et al. (1990), Rasio \& Shapiro (1995), Mochkovitch,
Guerrero \& Segretain (1997)) that if two CO WDs merge, the less massive
component disrupts into an inflated disk around its companion. In this
situation it has been argued that, due to a high effective accretion rate
from the disk, an off-center carbon ignition occurs and the more massive star
will be converted into an oxygen-neon-magnesium (ONeMg) WD
(Nomoto \& Iben (1985), Saio \& Nomoto (1985), Kawai, Saio \& Nomoto (1987),
Timmes, Woosley \& Taam (1994), Saio \& Nomoto (1998)). These discussions
do not take into account many of the complexities involved in the merger
process (see, e.g., Iben (1988)).
Perhaps one of the largest uncertainties plaguing the one dimensional
simulations which attempt to address the problem is the unknown rate at
which angular momentum diffuses out of the thick disk, allowing matter from
the disk to settle onto the white dwarf and adopt a structure describable
by a traditional one-dimensional model.

In this paper, we first present results of numerical experiments in which
CO-rich matter is accreted radially by an initially ``cooled-down'' CO WD of
0.8 \msun at rates in the range $10^{-5} - 10^{-8}$\msun\yrm1. 
We explore in detail the thermal response of the accreting structure and
point out the main physical parameters which drive the accretion process
and determine the final outcome. These experiments provide a reference
point for understanding results of further computations in which the
lifting effect of rotation is taken into account. 
In choosing an initially cooled-down structure, we ignore the possibility
that viscous heating prior to merger may be significant, but this is legitimate
since there is at present no theoretical or experimental framework for
estimating the magnitude of the relevant turbulent viscosity.

The plan of the work is as follows: 
in \S 2 we present the input physics and the adopted numerical techniques;
in \S 3 we discuss the evolution of the accreting structure as a function
of the accretion rate for standard (non-rotating) models;
in \S4 the lifting effect of rotation on the evolution of accreting models
is discussed.
Conclusions are summarized in \S5.

\section{Input Physics}

The accretion experiments have been computed with an updated version of the 
FRANEC code (Chieffi \& Straniero 1989), the main differences being discussed 
by Cassisi, Iben \& Tornamb\'e (1998). The equation of state is that 
computed by 
Straniero (1988) with succeeding upgrades; for the radiative low temperature 
opacities (T $< $12000 K) we adopt the tables provided 
by Alexander \& Fergusson (1994) whereas for higher temperatures the tables
provided by Huebner et al. (1977) are used. The contribution of conductive
opacities has been included following the prescriptions by
Mitake, Ichimaru \& Itoh (1984) and Itoh, Hayashi, \& Kohyama (1993).
The accretion process has been computed assuming that the accreted matter
has the same specific entropy as the 
matter at the surface of the WD (for a discussion of this topic see Taam 
(1980a,b), Iben (1982), Nomoto \& Iben (1985)). We have adopted
values of the accretion rate in the range $10^{-8} \div 10^{-5}$ \msun\yrm1; 
the upper limit is an estimate of the maximum accretion rate at which the 
transferred matter is accepted by the accreting WD (Eddington limit), whereas 
the lower limit has been
fixed so that the accretion process can produce a SN Ia event well
before the onset of crystallization in the inner zones of the WD. The 
chemical composition of the accreted matter has been chosen to be equal to the 
chemical composition of the matter in the surface layers of the CO WD, namely
$X_{^{12}C}=0.507$, $X_{^{16}O}=0.482$ and $X_{^{20}Ne}=0.011$.

\placefigure{fig1}

\section{Accretion with Standard Conditions}

\subsection{The Initial Model}

The initial model is a 0.8 \msun CO WD obtained by evolving a solar metallicity
pure He star from the helium main sequence up to the cooling sequence
(for more detail see Limongi \& Tornamb\'e (1991)). In the computation
of the central helium-burning stage,
semiconvection has been accounted for but breathing pulses at the border of the
convective helium core have been neglected, as suggested by the results of 
Caputo et al. (1989).

When the model reaches the cooling sequence and the  helium-burning shell
switches off, the oxygen profile in the core exhibits a maximum and
the carbon profile exhibits a minimum (see {\em panel a} in Fig. 1).
This particular feature forms when, after the exhaustion of helium at the
center, the radiative helium-burning shell goes through the 
region where helium was partially depleted during the outward extension of the 
convective helium core in the central helium-burning stage; in this zone,
due to the reduction in the local helium abundance,
$\alpha$-captures on ${^{12}{\rm C}}$ compete with 3$\alpha$ reactions.
The extension in mass of this oxygen-enriched region depends on the
efficiency of semiconvection, while the level of the peak depends on the
adopted cross section for the ${^{12}{\rm C}}(\alpha,\gamma){^{16}{\rm O}}$
reaction. Adopting the prescriptions in Salaris et al. (1997), the
chemical profiles have been modified as in {\em panel b} of Fig. 1 and
then the cooling phase has been computed. The physical characteristics
of the cool model are:
$\log(L/L_\odot)=-2.858,\ \log(T_{\rm eff})=4.049,
\ \log(\rho_{\rm c})=7.036, \ {\rm and}\ \log(T_{\rm c})=7.015$. 

\subsection{Time Scales of Relevance for an Accreting White Dwarf}

The thermal evolution of a CO WD accreting CO-rich matter can be discussed 
in terms of several relevant time scales. For example, the
final outcome of the accretion process can be understood by examining the
behavior in time of the ratio of a thermal diffusion time scale
(Henyey \& L'Ecuyer 1969) and a compressional heating time scale
(Sugimoto \& Nomoto 1977, Nomoto 1982a)

Ignoring all but gravothermal energy generation, conservation of energy
requires that
\begin{equation}
         \rho\left({dU\over dt}+P{dV\over dt}\right)=-\nabla\cdot {\bf F}.
\end{equation}
On the left hand side of this equation,
$U$ is the thermal energy per unit mass, $P$ is the pressure, $V$ is the
specific volume ($V=1/\rho$), $\rho$ is the density, and derivatives with
respect to time are co-moving derivatives in Lagrangian coordinates. On the
right hand side, assuming that energy flow is by radiation and conduction only,
\begin{equation}
  {\bf F} = -{16\sigma\over 3} {T^3\over \kappa \rho} {dT\over dr} \hat{\bf r},
\end{equation}
where $T$ is the temperature, $\sigma$ is the Stefan-Boltzmann constant, 
$\kappa$ is the opacity, and $\hat{\bf r}$ is a unit vector in the radial
direction.

A local compressional heating time scale can be defined by 
setting the left hand side of eq. (1) equal to zero, giving
\begin{equation}
        {1\over \tau_{\rm comp}}
              = \left({1\over U}{dU\over dt}\right)_{\rm adiabatic} 
                 = \left({P\over \rho U}\right) {d\ {\rm ln} \rho\over dt}
\end{equation}
Since the quantity $\displaystyle{\left({P\over \rho U}\right)}$ is nearly
constant, varying in value from 1 (electrons with non-relativistic velocities)
to $\displaystyle{{2\over 3}}$ (relativistically degenerate electrons),
it is reasonable to adopt the simpler expression
$\displaystyle{{1\over \tau_{\rm comp}} = {d\ {\rm ln} \rho\over dt}}$.

Nomoto \& Sugimoto (1977) set
\begin{equation}
   {1\over \tau_{\rm comp}}= {d\ {\rm ln} \rho\over dt}=
         \left({\partial\ {\rm ln} \rho\over \partial t}\right)_q
                  -\left({\partial\ {\rm ln} \rho\over \partial {\rm ln}
M_r}\right)_t
                     {d {\rm ln} M\over dt},
\end{equation}
where $M$ is the total mass, $M_r$ is the mass
at radius $r$, and differentiation is with respect to mass fraction
$\displaystyle{q={M_r\over M}}$ and time $t$ as independent variables.
Neglecting the first term in eq. (4), Nomoto \& Sugimoto adopt
\begin{equation}
   {1\over \tau_{\rm comp}} \sim 
                  -\left({\partial\ {\rm ln} \rho\over \partial {\rm ln}
M_r}\right)_t
                     {d\ {\rm ln} M\over dt}
                           ={GM_r^2\over 4\pi r^4P} 
         \left({\partial\ {\rm ln}\rho\over \partial\ {\rm ln} P}\right)_t
                      {1\over \tau_{\rm acc}},
\end{equation}
where
\begin{equation}
                \tau_{\rm acc}=\left({d\ {\rm ln} M\over dt}\right)^{-1}
\end{equation}
is the time scale for accretion.

Differentiating with respect to $q$ and $M$ as independent variables,
Nomoto (1982a) gives
\begin{equation}
   {1\over \tau_{\rm comp}}= \left({d\ {\rm ln} \rho\over dt}\right)_{{M_r}}=
         \left[\left({\partial\ {\rm ln} \rho\over \partial {\rm ln}
M}\right)_q
                  -\left({\partial\ {\rm ln} \rho\over \partial {\rm ln}
q}\right)_M
                     \right] {1\over \tau_{\rm acc}},
\end{equation}
showing that, in all approximations, $\tau_{\rm comp}$ is proportional to
$\tau_{\rm acc}$.

A local thermal diffusion time scale can be defined as the 
time required by a thermal signal to propagate over a specified distance.
Equations (1) and (2) give
\begin{equation}
         \rho\left({dU\over dt}+P{dV\over dt}\right)=
      \rho C_P {dT\over dt}-
    \left({\partial\ {\rm ln} \rho\over \partial\ {\rm ln} P}\right)_T
             {dP\over dt} =
                            \left({d\over dr} + {2\over r}\right)
   \left({16\sigma\over 3} {T^3\over \kappa \rho} {dT\over dr}\right),
\end{equation}
where $C_P$ is the specific heat at constant pressure. If we set
$\displaystyle{{dP\over dt} = 0}$, neglect $\displaystyle{{2\over r}}$
relative to $\displaystyle{{d\over dr}}$, and assume that
$\displaystyle{{T^3\over \kappa \rho}}$ varies much less rapidly with 
distance than $\displaystyle{{dT\over dr}}$, then
\begin{equation}
         {d T\over d t} \sim D {d^2 T\over d r^2},
\end{equation}
where
\begin{equation}
      D= K^2 = {16\sigma\over 3} {T^3\over \kappa C_P \rho^2}.
\end{equation}

Equation (9) is the well known equation for the diffusion of heat
and its solution for many situations is proportional to (e.g., Margenau
\& Murphy 1943)
\begin{equation}
          \exp{\left(-{\delta r^2\over 4 D t}\right)},
\end{equation}
where $\delta r$ is the distance from an initial delta-function perturbation
in $T$. This motivates the choice of
\begin{equation}
            \tau_{{\rm diff},\delta r} = {\delta r^2\over 4 D}
\end{equation}
as a measure of, or ``time scale'' for, the diffusion of heat across the
distance $\delta r$.

As suggested by Sugimoto (1970), an appropriate length scale is the
local pressure scale height $H_P$, leading to a ``local'' diffusion
time scale
\begin{equation}
            \tau_{\rm diff,loc} = {H_P^2\over 4 D},
\end{equation}
(Sugimoto chooses a value 4 times larger than this). One might naively
assume that a time scale for the propagation of heat over many
scale heights is given by the sum of local time scales over these scale
heights.  However, as discussed by Henyey \& L'Ecuyer (1969), the appropriate
approximation is instead
\begin{equation}
\tau_{\rm diff,glob}= \left(\int_{{r}_1}^{{r_2}}{{1\over 2 K} dr}\right)^2
         = \frac{3}{64\sigma}\left[\int_{{r_1}}^{{r_2}}
      \left(\frac{\kappa C_p}{T^3}\right)^{1/2} \rho dr\right]^2,
\end{equation}
where the integral extends from $r_1$ to $r_2$.

In the following, we will make use of a ``global'' compressional heating
timescale defined as $\tau_{\rm ch}=\tau_{\rm acc}$ and of a ``global''
diffusion time scale $\tau_{\rm diff}$ equal
to the value of $\tau_{\rm diff,glob}$ given by eq. (14) when $r_1 = 0$ and
$r_2 = R$, where $R$ is the radius of the star.

The gravitational energy released by accreted matter is partly stored
in layers near the surface as thermal energy, but partly diffuses inward,
increasing the thermal content of layers ever deeper below the surface.
If \tdf $>$ \tchend, heating of layers near the surface may lead to the
off-center ignition of carbon, well before the white dwarf reaches the
Chandrasekhar limit (Nomoto 1982a,b, Nomoto \& Iben 1985).
If \tdf $<$ \tchend, energy can be transferred inward efficiently enough
that carbon ignition occurs at the center or, if oxygen has replaced
carbon at the center, at the base of a carbon-rich layer
(Isern et al. 1983, Hernanz et al. 1988).

Choosing $D=R_{\rm star}\sim 10^{-2} R_\odot$, $T=T_{\rm c}\sim 10^7$ K
(isothermal structure), $C_p \sim 10^7$ erg/g K, $\rho$ equal to the average
density of the structure ($\sim 10^6$ g/cm$^3$), and $\kappa$ equal to
the mean conductive opacity ($\kappa\sim 10^{-4}$ ${\rm cm}^2$/g), our
initial model is characterized by \tdf $\sim 3\times 10^7$ yr. Insisting
that \tdf = \tchend, the maximum value for the accretion rate suitable
for the efficient transfer toward the inner zones of gravitational energy
released by accretion can be estimated as
$\dot{M}_{\rm max}\sim 3\times 10^{-8}$\msun\yrm1. It follows that,
if \mdot $< \dot{M}_{\rm max}$, the thermal energy excess in the outer
layers is transferred to the interior and carbon burning is ignited
at the center. 

The value of \tdf just estimated refers to the initial model. Since \tdf
actually evolves in time as a consequence of the evolution of the physical
conditions in the accreting structure, it is necessary to examine its
value in the evolving models. 

\placefigure{fig2}

In Fig. 2, the evolution of \tdf (solid lines) is shown in comparison with
the evolution of \tch (dashed lines) for all of the models computed in the
present work, as listed in Table 1. It is evident that,
for \mdot $<$ \mdotend$_{\rm max}$ ({\em panel i}), the mass of the accreting
WD reaches the Chandrasekhar limit, so that central carbon burning occurs.
It is apparent also that, for the cases with 
\mdotend$_{\rm max}$ $<$ \mdot $<$ $10^{-6}$ \msun\yrm1
({\em panels e} to {\em h}) ,
the gravitational energy deposited in outer layers is transferred efficiently
toward the center since, during the accretion process, the 
compressional heating time scale increases while the thermal diffusion time 
scale decreases until the ratio of the two time scales is reversed. This
occurrence can be easily explained by observing that, on increasing the
total mass, the radius and the conductive 
opacity of the accreting structure decrease while density and temperature 
increase. This behavior is common to the entire set of computed models, 
but if $\dot{M} > 10^{-6}$ \msun\yrm1 , carbon-burning ignition occurs in
the outer layers well before a significant amount of the deposited
gravitational energy can be transferred toward the center.
Fig. 2 discloses further that, for small enough accretion rates (say,
less than $5\times 10^{-7}$\msun\yrm1), the thermal diffusion time scale
profile has two different slopes, the inflection point corresponding to
the time when the thermal wave reaches the center. 

\placefigure{fig3}

This behavior can be better appreciated in Fig. 3 where the evolution of
the mass coordinates of the maximum temperature (${M}_{T-{\rm max}}$)
and of the central temperature ($T_{\rm c}$) are reported
together with \tdf as functions of the total mass of the accreting 
structure for the case \mdot $=5\times 10^{-8}$ \msun\yrm1. Where the
central temperature increases most rapidly with mass (mass coordinate
$\sim 1.05 \pm 0.05$ \msunend),
the rate of change in the slope of \tdf decreases rapidly, becoming almost
flat. The maximum in temperature does not reach the center until the whole
structure has become isothermal. Thereafter, the evolution of the central
temperature is driven by homologous compression.

The final sudden increase of the central temperature and 
the corresponding decrease in \tdf when the total mass of the 
accreting WD reaches $\sim 1.35$ \msun is a consequence of the strong
compression due to the approach to the Chandrasekhar mass. 

A discussion of global behavior can also be conducted in terms of local time
scales. By way of illustration, we examine the cases of accretion at rates
marking the boundaries of the region we have explored, namely, 
$10^{-5}$ \msun \yrm1 and $10^{-8}$ \msun \yrm1.
In Fig. 4, the evolution of the temperature, the local compressional
heating time scale \tchloc given by eq. (2), and of two ratios involving
\tchloc, \tdfloc given by eq. (11), and $\tau_{\rm evo}$, the time
elapsed since the beginning of the experiment, are shown for the case
$10^{-5}$ \msun \yrm1.

\placefigure{fig4}

In the inner regions ($M < 0.65$ \msunend), the heat diffusion time
scale is long compared with the compressional heating time scale,
so the increase in temperature is due entirely to local compressional
heating. In central regions ($0.65 < M < 0.8$ \msunend), the heat
diffusion time scale decreases in the outward direction to become
comparable to the compressional heating time scale, and heating is due to
a combination of compressional heating and inward diffusion, the contribution
of diffusion increasing outward with respect to the contribution of
compression. In the outermost accreted layers ($M > 0.8$ \msunend),
\tdfloc is everywhere smaller than \tchloc and, since also everywhere \tdfloc
$< \tau_{\rm evo}$, diffusion efficiently redistributes energy liberated 
by compression both inward and outward. The peaks in the temperature profiles
show that the ``watershed'' marking the place where the heat flow changes
directions is approximately at the center of the accreted layer.

Figure 5 shows the evolution of the same quantities displayed in Fig. 4,
but for the case $10^{-8}$ \msun \yrm1. In this case, \tdfloc is always 
larger than \tchloc everywhere in the interior; as \tevo increases,
compressional energy released everywhere is redistributed in such a way
that the whole structure becomes nearly isothermal.

It is worth pointing out that, over the matter in the initial white dwarf,
the mass averaged value of $\tau_{\rm ch}$ is essentially $\tau_{\rm acc}$,
whereas, in the accreted layer, $\tau_{\rm ch}$ is everywhere much less
than $\tau_{\rm acc}$. So, what we really do when we compare
$\tau_{\rm diff,glob}$ with $\tau_{\rm acc}$ is compare the time it takes
energy deposited in accreted layers to make its way into the deep
interior of the original white dwarf with the time it would take
these layers to heat up without inward diffusion.

\placefigure{fig5}

\subsection{The Accretion Experiments}

In Fig. 6 we plot the evolution of the accreting structures in the HR diagram
while in Fig. 7 we report the evolution of the surface radius as a function 
of the total mass for all the models listed in Table 1. 
It is evident that, at the beginning of the accretion process, the
structure returns along the cooling sequence discussed in \S 3.1 and
expands since the
gravitational energy deposited at the surface causes a local increase in
temperature and, hence, in luminosity. It is important to keep in mind that
the initial WD is a very compressed isothermal structure so that the local
efficiency of the inward thermal diffusion is very small (\tdf is very large);
indeed, all of the energy delivered by the accreted matter is locally
stored as thermal energy. The increase in pressure related to the increase
in temperature leads to a rapid expansion of the outer layers; the expansion 
induces a decrease in the local density while the continuous deposition
of matter produces an increase in the local temperature. In fact, the value
of \tdf in surface layers begins to decrease and, when the energy excess
can be transferred to the underlying zones, that is,
when the local \tdf becomes less than \tchend, the 
structure begins to contract. As is clearly shown in Figs. 6 and 7, the
larger the accretion rate, the larger is the radius at the end of the
expansion phase and the larger are the luminosity and the effective
temperature; as the accretion rate is increased, the rate at which
gravitational energy is deposited in surface layers is increased and,
hence, the induced expansion becomes larger.

\placefigure{fig6}

\placefigure{fig7}

In  Fig. 8, we plot the evolution of the center of the accreting WD 
in the $\rho-T$ plane and in Table 1 we report for each adopted value of the 
accretion rate the final mass (in \msunend ), the mass coordinate
(in \msunend) at which carbon burning is ignited, and the value of
temperature (in K) and density (in g cm$^{-3}$) at this point. Only for 
\mdot smaller than $10^{-6}$ \msun\yrm1 does central carbon ignition occur.
Three different phases can be distinguished in the thermal evolution of models
which ignite carbon burning at the center:

\placefigure{fig8}

\begin{itemize}
\item the first phase, during which the thermal evolution of the center 
is determined by homologous compression. In this phase, the increase in
mass of the WD induces the compression of the whole structure while
the gravitational energy delivered by the accreted matter is stored in the
outermost layers;
\item the second phase, characterized by a rapid increase in the central
temperature. This rapid increase in central temperature is a consequence
of the fact that the
gravitational energy initially localized in outer layers has diffused to
the center. During this phase, homologous compression continues to be
active and drives the evolution of the central density. This phase is
concluded when the structure becomes almost isothermal;
\item the third phase, during which thermal diffusion and homologous
compression are both effective. At this stage, the structure is
isothermal, as it was when the accretion process began, but now the
temperature level is higher; as a consequence, gravitational energy is not
stored for long in outer layers but is almost immediately transferred
to underlying zones (see the drop in \tdf in Fig. 3 from the first to the
third phases). In addition, the gravitational energy delivered by the accreted
matter is negligible with respect to the heating produced by homologous
compression, the efficiency of which is independent of the thermal content 
of the accreting structure; as a consequence, during this phase, homologous 
compression becomes the leading factor which governs the increase in the
interior temperature.
\end{itemize}

It is interesting that the rate of increase in the central temperature in 
the third phase is smaller than that in the first one, whereas one might
reasonably expect that the combined effect of both thermal
diffusion and homologous compression would produce a more rapid 
increase in the central temperature during the third phase. The
actual behavior may be explained by observing that the rate of energy
loss via plasma neutrino emission increases with increasing temperature,
thus leading to a reduced efficiency of heating of the central regions
where temperatures are highest.

In Fig. 9 are plotted, for every adopted accretion rate, temperature profiles 
at different times during the accretion process.
For models with accretion rates larger than $10^{-6}$ \msun\yrm1, the 
evolution of both temperature and density at the center is determined only by 
homologous compression. In fact, for such models, the thermal excess introduced
by the accretion process into outer layers does not reach the center before 
runaway carbon burning (initiated at the point where the temperature is at
a maximum) terminates the process. The model accreting matter at
\mdot $=2\times 10^{-6}$ \msun\yrm1 defines a transition line: the
physical conditions suitable for carbon burning are achieved first near
the center (see panel d) in Fig. 9, M = 0.2 \msunend), but
quite soon also in outer layers where the maximum in temperature is located
($M=1.375$ \msunend). As time goes on, however, the rate of delivery of
nuclear energy increases more quickly in the outer layers than in the
deep interior and the thermonuclear runaway occurs first in the outer layers.

\placefigure{fig9}

In Fig. 10 are plotted the density profiles for the same models described 
in Fig. 9. The initial expansion phase is 
clearly shown by the decrease in the surface density whereas the action of the
homologous compression is evidenced by the self-similar evolution of the
density profiles. The same behavior is evident in the evolution of the
temperature profiles for the case $\dot{M}=10^{-5}$ \msun\yrm1; as time
increases, the temperature level of the inner regions increases homologously
while the continuous accretion process induces the formation of a peak in
temperature in the outer layers. 

\placefigure{fig10}

Models with $10^{-7} < \dot{\rm M} \le 10^{-6}$\msun\yrm1 ({\em panels e} and 
{\em f} in Fig. 9) achieve the conditions for carbon ignition in the inner
zone when the maximum in temperature is still located in outer layers.
This is also apparent from an inspection of Fig. 11, where we report
profiles in the $\rho-T$ plane for the models which have achieved ignition
conditions. This behavior is due to the fact that the energy loss rate via
plasma neutrinos, $\varepsilon_{{\rm plasma}-\nu} \propto T^3$, is
very large near the mass coordinate where the maximum in temperature is
located. Hence, the gravitational energy delivered by the accreted matter
causes a steepening of the temperature gradient and, paradoxically, the
energy-loss sink due to $\varepsilon_{{\rm plasma}-\nu}$ becomes deeper,
reducing the amount of heat which can be transferred toward the center.
Due to the effect of plasma neutrino losses, there is no carbon ignition
for densities in the range $10^{7} - 10^{9}$ g/cm$^3$.

\placefigure{fig11}

Models with accretion rates smaller than $\dot{\rm M} \le 10^{-7}$
\msun\yrm1 succeed in removing the energy excess deposited in
outer layers because the gravitational energy is deposited more 
slowly than it diffuses inward. The surface energy reservoir is smaller
and, hence, the energy losses via plasma neutrinos are smaller, so that
the heat transfer inward is very efficient. In addition, the homologous
compression induced by the accretion process is less rapid and the
accreting star approaches the Chandrasekhar mass without carbon
burning being ignited at a significant level anywhere within the
structure. On the other hand, as the total mass approaches \mch,
the rate of contraction becomes larger and the effect of the induced
compression, which is larger at the center, determines the physical
conditions suitable for a central carbon ignition.
During subsequent evolution under
highly degenerate physical conditions, the energy delivered by carbon burning
is stored locally, increasing the local temperature; the onset of convection 
(see {\em panels f} to {\em i} in Fig. 9) marks the onset of the thermonuclear
runaway which will cause the disruption of the accreting star as a
Type Ia Supernova. Although in these models the leading parameter in the
evolution up to the carbon-ignition stage is the compression induced by
strong contraction, the matter deposited on the surface plays a 
not completely negligible role in determining the ignition conditions.
On decreasing the accretion rate, the density at which carbon ignition occurs
becomes larger and, correspondingly, the temperature becomes smaller.

\section{The Effect of Rotation}

\subsection{Physical Assumptions and Numerical Procedures}

To explore the effect of stellar rotation on the thermal response of CO WDs
accreting matter we have included the lifting effect into the 
hydrostatic equilibrium equation according to the prescriptions in 
Dominguez et al. (1996):
\begin{equation}
\frac{dP}{dr}=-\frac{G M}{R^2}\rho \left(1-f\right),
\end{equation}
where $f$ is the ratio between the centrifugal and gravitational
forces $\displaystyle{f={F_c\over F_g} = {\omega^2 r\over g}}$. In order
to maintain the 
spherical symmetry in the treatment of the problem, the centrifugal force 
on a generic mass element $dm$ is averaged over a sphere of radius $r$ 
so that its value is provided by the following relation:
\begin{equation}
< F_c >=\frac{\int_{S_r} F_c dS_r}{S_r}=dm\frac{\int_0^\pi 2\pi\omega^2 r^3
\sin^2\theta d\theta}{4\pi r^2}=\frac{\pi}{4}dm\omega^2 r.
\end{equation}

This treatment is equivalent to those by Kippenhahn \& Thomas (1970) and
Endal \& Sofia (1976, 1978) except that we neglect the corrective factor
included by these authors in the expression for the radiative gradient
(see Dominguez et al. 1996). Since WDs are chemically fairly homogeneous and
are very compact objects, the efficiency of angular momentum redistribution
may be large enough (Maeder \& Meynet 2000) that the accreting WD can be
treated as a rigid rotator, and we assume this to be the case.

As discussed by Tassoul (1978), a rotating fluid undergoes instabilities
which must be accounted for in a realistic description of the phenomenon.
By adopting a rigid-rotator model, neither the Solberg nor the
Goldreich-Schubert-Fricke instabilities arise. On the other hand, rotation
could induce deformations of the equipotential surfaces in a such a way that 
meridional circulations on a large scale arise (von Zeipel paradox).
In this case, deviations from spherical symmetry can be checked by computing
a dimensionless quantity $\gamma$ defined as the ratio between rotational
and gravitational energy ($\gamma=E_{\rm rot}/E_{\rm g}$).
If $\gamma > 0.14$,
the equilibrium configuration is no longer spherical and meridional
circulations cannot be neglected. 
In our computations, the $\gamma$ parameter is always smaller than 0.1, so 
meridional circulations have not been taken into account. 
Finally, if the centrifugal force acting on a mass element $dm$ 
becomes larger than the gravitational force ($f>1$ in eq. 3), this element 
will experience a radial acceleration outward, achieving a velocity greater
than the escape velocity (Roche instability) and the accretion process 
will be terminated. The angular velocity for which $f=1$ is defined 
as the {\em critical angular velocity}\ $\omega_{\rm crit}$, and its
value is given by
\begin{equation}
       \omega_{\rm crit}=\sqrt{g/r},
\end{equation}
where $r$ is the distance from the center to the
mass element $dm$ and $g$ is the local value of the gravitational acceleration.
This instability has been checked in detail during the evolution of the 
accreting models and will be addressed in the discussion of results.

In the present computation, it is assumed that the accreted matter flows
to the WD from a co-rotating accretion disk and that the deposited material 
has the specific angular momentum $j_{\rm D}=\omega d^2$, where $d$
is the radius of the accreting WD and $\omega$ is the instantaneous angular
velocity of the WD. The rate of change of the angular velocity
$\omega$ of the WD model is then given by $\displaystyle{{d\omega\over dt}
 = {j\over I} \dot{M}}$, where $I$ is the moment of inertia of the model.

The initial angular velocity, and, as a consequence, the initial total
angular momentum of the accreting WD have been fixed parametrically by
assuming that $\omega = \alpha\ \omega_{\rm crit,0}$, where
$\omega_{\rm crit,0}$ is the initial value of $\omega_{\rm crit}$
and $\alpha$ is a constant ($< 1$).
In our case, $\omega_{\rm crit,0} = 0.625$ rad/s and, in the computations,
we have adopted $\alpha$ = 0.1, 0.3, 0.5, 0.7, and 0.9.

\subsection{Numerical Results}

The inclusion of rotation in the hydrostatic equilibrium equation has a
``lifting'' effect, reducing the local value of pressure throughout
the structure in such a way that the accreting star is both less dense 
and cooler than in the standard case (rotation neglected). 
While the density is only moderately decreased, the decrease in temperature
throughout the accreting structure leads to a large enough increase in the
thermal diffusion time scale to have a noticeable effect
on the evolutionary behavior of the model. This is clearly
shown in Fig. 12 where the initial value of \tdf is plotted as a function
of the initial angular velocity. 

\placefigure{fig12}

On the other hand, the effect of compression induced by the accretion process
is only slightly modified by the inclusion of rotation. Since \tdf is far
more sensitive to the inclusion of rotation than is \tchend, an increase in
the initial angular velocity of the WD always means a decrease in the value
of the accretion rate below which central carbon ignition 
(as defined in \S 3.2) will occur.
Evolution during the expansion phase at the beginning of the accretion
process is modified in a similar way. Since the initial expansion 
process is totally driven by the ratio \tdfend/\tchend,
for a fixed accretion rate (that is, for a given amount of 
gravitational energy delivered onto the WD per unit of time), the local energy 
excess in surface layers and, hence, the induced expansion become larger,
the larger the initial angular velocity. This dependence is illustrated
in Fig. 13, where the evolution of surface radius versus total mass is
shown as a function of initial angular velocity.

\placefigure{fig13}

When rotation is not taken into account, the expansion comes to a halt
due to a decrease in \tdf in surface layers; gravitational energy is
transferred efficiently to underlying zones from surface layers, which
consequently begin to contract (see the discussion in \S 3). When rotation
is taken into account, the Roche instability can occur, leading to an
expansion phase during which $\omega_{\rm crit}$ decreases in layers
near the surface.
In this case, the accreting structure may become gravitationally unbound
after only a small amount of matter has been accreted.
In Fig. 14 are shown the time evolution of several quantities
for the model with $\dot{M}=10^{-8}$\msun\yrm1 and 
\o0 =0.5 $\omega_{\rm crit,0}$. The quantities are the thermal diffusion
time scale near the surface (panel a), the surface temperature (panel b),
the radius (panel c). Panel d shows the evolution of the WD angular velocity
$\omega$ (lower curve) and of the critical
angular velocity in the outermost layers $\omega_{\rm crit,s}$.
As the surface radius increases, 
$\omega_{\rm crit,s}$ decreases while,
because of the deposition of angular momentum by accreted matter,
$\omega$ increases almost linearly with increasing WD mass.

\placefigure{fig14}

From these results, it follows that, for each value of \mdotend, there
exists a threshold value \oth for $\omega_{0}$ such that,
if $\omega_0 > \omega_{th}$,
the accreting structure becomes gravitationally unbound due to the expansion
during the first phase of the accretion process. In the $\dot{M} - \omega_0$ 
plane of Fig. 15, the shaded region gives the initial angular velocity
and accretion rates for which the accreting WD becomes gravitationally
unbound when its total mass approaches 1.4 \msunend. 

\placefigure{fig15}

Models which survive the expansion during the first phase of the accretion
process recontract since the decrease in \tdf in outer layers leads
to an efficient transfer inward of the gravitational energy released.
During this phase, the evolution of the accreting
structure resembles the evolution of a non-rotating model, although,
for a given value of the accretion rate, the whole structure is cooler and
less dense for larger \o0. In Fig. 16, we show the same quantities as in
Fig. 14 for a case in which the accretion rate is the same
(\mdot $=10^{-8}$\msun\yrm1), but in which the initial angular velocity
is smaller ($\omega_0/\omega_{\rm crit,0}$ = 0.1). In this case,
the initial expansion is not large enough to significantly slow the
decrease in \tdfend, and both $\omega$ and $\omega_{\rm crit,s}$ increase
steadily with increasing WD mass as
the energy excess deposited in outer layers is efficiently
transferred toward the center, as is clearly depicted in Fig. 17 where 
temperature profiles for various models during the accretion process are
shown. 

\placefigure{fig16}

As detailed in Fig. 16, the accreting WD spins up as its radius increases,
with both $\omega$ and $\omega_{\rm crit,s}$ increasing steadily; as a result,
the temperature level throughout the WD increases homologously (see also
Fig. 17). When the WD mass reaches $\sim 1.3$ \msunend, the centrifugal
force begins to compete with the gravitational force; thereafter,
the increase in the total angular momentum of the structure causes the
surface layers of the WD to expand and the compression induced by the
accreted matter is no longer able to balance the radiative losses. As a
consequence, the entire structure cools down (see the last profile in Fig. 17)!
Notice further that the expansion produces a decrease in $\omega_{\rm crit}$
to the extent that the accreting structure becomes gravitationally unbound;
the accretion process comes to a halt. 

\placefigure{fig17}

It thus appears that the occurrence of the Roche instability prevents the
accreting models from achieving the physical conditions necessary for the
ignition of carbon; the accreting structures are not only too
cool but, in addition, the centrifugal force due to rotation prevents the
contraction in interior regions which produce heating by the conversion of
gravitational potential energy.

\subsection{The Dependence of the Final Outcome on $j_{\rm D}$}

The scenario summarized in Fig. 15 depends on the amount of angular 
momentum deposited by the accreted matter: in fact, the increase in the
angular velocity, particularly during the early phase of the accretion
process, is totally controlled by the deposition of angular momentum.
In order to make this point clearer, we have re-computed the
previously discussed models assuming that the accreted matter flows to
the WD from an accretion disk at rest ($j_{\rm D}=0$). In Fig. 18,
we report the evolution of the surface radius for models accreting
CO-rich matter at the rate \mdot $=10^{-7}$ \msun \yrm1 with
different initial angular velocities, assuming that the accreted matter 
deposits (dashed lines) or does not deposit (solid lines) angular
momentum. It is clear that, if angular momentum is not added, the
accreting structure with initial angular velocity smaller than
0.9 $\omega_{\rm crit,0}$ survives the initial expansion and thereafter
contracts as a consequence of the increase in the total mass. In addition,
if $j_{\rm D}=0$, for \o0 in the range $0.3\div 0.9$ $\omega_{\rm crit,0}$,
the accreting structure ignites carbon burning at the center since the
total mass is larger than 1.4 \msunend . It is also clear that, the
larger the initial value of the angular velocity, the larger is the
final total mass of the accreting WD.
This occurrence is consistent with the known fact that the
Chandrasekhar mass for rotating structure is larger than that
for non rotating models, depending strongly on the angular velocity
(see Anand 1968, Ostriker and Bodenheimer 1968).

\placefigure{fig18}

By assuming that the accreted matter does not deposit angular momentum,
the morphology of the \mdot-\o0 plane changes from that shown in Fig. 15
to that shown in Fig. 19. The changes are worth emphasizing:

\begin{itemize}
\item for large values of both \o0 and \mdot (the region labeled 
``$M_f<1$ \msunend''), the accreting models experience the Roche
instability when only a small amount of matter has been accreted
($\Delta M_{\rm acc}\le 0.1$ \msunend ). 
This zone corresponds to the non-shaded region in Fig. 15, so that
its lower boundary represents the \oth previously defined; 
\item for large values of \o0 and small values of \mdot (the region 
labeled ``$\varepsilon_{\rm n} < \varepsilon_\nu$''), the accreting
structures never achieve the physical conditions necessary for carbon
ignition, although they do not become gravitationally unbound. The
computation of these models has been halted because the adopted
input physics, in particular the equation of state, is inadequate.
In any case, the final mass of the accreting structure is larger
than 1.4 \msunend ;
\item for small values of both \o0 and \mdot (the region labeled
``Central Ignition''), the accreting structures ignite carbon burning
at the center when the total mass approaches (and in some case exceeds)
the non-rotating Chandrasekhar mass;
\item the zone labeled ``Central and Off-Center C-ignition'' represents 
structures which ignite carbon burning both at the center and in the
outer layers. This zone marks the transition between models which
become gravitationally unbound and models which succeed in igniting 
carbon burning;
\item models in the region above the dashed line and below the present
\oth line become gravitationally unbound when the total 
mass of the accreting structure approaches the non-rotating
Chandrasekhar mass. This region corresponds to the 
shaded zone in Fig. 15.
\end{itemize}

\placefigure{fig19}

It is evident that, on including the lifting effect of rotation on the
evolution of a CO white dwarf accreting CO-rich matter, the zone in the
$M_{\rm WD}$ - \mdot plane suitable for a central 
carbon ignition becomes narrower. If matter is accreted
at rest, carbon burning is ignited at the center for the same
values of the accretion rate as in the standard case 
(\mdot $< 10^{-6}$\msun\yrm1),
but only for small values of the initial angular velocity (see Fig. 19), 
whereas, if the matter is accreted from a co-rotating disk,
the WD never ignites carbon burning. In this latter case,
the increase of the angular velocity due to the deposition of
angular momentum by the accreted matter induces the expansion of
the accreting structure which becomes gravitationally unbound.

\subsection{The Braking Phase}

Interestingly enough, rotating models with small values of
both the initial $\omega$ (namely \o0 $<$ 0.3 $\omega_{\rm crit,0}$) and 
accretion rate (namely \mdot $< 10^{-7}$ \msun\yrm1 ) undergo the Roche
instability when their total mass is of the order of the non-rotating 
Chandrasekhar mass. In this case, the ratio of rotational 
and gravitational energies ($\gamma=E_{\rm rot}/E_{\rm g}$)
approaches the critical value $\sim 0.14$
and, as a consequence, the accreting structure has to deform into
an ellipsoidal shape (see \S 4.1); due to the induced deformation,
the mass distribution of the white dwarf becomes anisotropic in such
a way that gravitational wave radiation is emitted (see Shapiro \& 
Teukolski 1983, 1990). The emitted waves carry away rotational energy
and angular momentum and the total angular momentum of the
braking structure evolves according to an exponential law $J=J_0
e^{-t/\tau}$, where $J_0$ is the value of the angular momentum at the
beginning of the slowing down phase and $\tau$ is the time scale on
which the process occurs; the latter quantity is determined by the
efficiency of braking and, therefore, by the angular velocity
of the structure when $\gamma$ approaches the critical value. Since
the intensity of the emitted GWR depends on $\omega^2$, we assume
that the braking process becomes the driving mechanism for the
evolution of the accreting white dwarf when the angular velocity 
is of the order of 1 rad s$^{-1}$.

It is worth noticing that, on decreasing the total angular momentum,
the effective gravity acting on a given mass element increases and,
as a consequence, the whole structure experiences a
contraction which produces the heating-up of the entire star. 
In addition, the contraction induced by the braking produces an
increase in the angular velocity of the white dwarf so that the
$\gamma$ parameter actually remains of the order of the critical value.
This means that the white dwarf continues to emit GWR copiously as it
contracts.

These considerations suggest that the braking process acts as a
compression mechanism and that the $\tau$ parameter in the
exponential braking law can be interpreted as a compressional
heating time scale; we can therefore argue
that the final outcome of the braking phase depends on the
ratio between $\tau$ and \tdfend, the thermal diffusion time scale
of the considered structure. 

In order to explore this conjecture, we have constructed an
evolutionary sequence which begins with a model at the end of
the previously explored sequence with initial
angular velocity $\omega_0=0.1 \omega_{\rm crit,0} = 0.067$ rad/s
and accretion rate \mdot $=10^{-8}$ \msun\yrm1 . In this last
model, $\gamma$ is of the order of the critical value,
the total mass of the white dwarf is $\sim 1.42$ \msun , the
angular velocity is $\sim 1.1$ rad/s, and the value of \tdf
is $\sim 2\times 10^7$ yr; at this
time the structure is isothermal, as shown in Fig. 17.

In order to explore the effect of GWR braking on further evolution,
we choose two values of $\tau$:
$\tau=10^7$ yr, which corresponds to a low braking efficiency, and
$\tau=10^3$ yr, which corresponds to a high braking efficiency. As
discussed in Shapiro \& Teukolski (1983) and in Ma (2002),
the GWR-driven braking process is not important on time scales
greater than $10^7$ yr since, then, the
slowing down is dominated by viscosity. 

In the first case ($\tau=10^7$ yr), the homologous compression
induced by GWR braking produces an increase in temperature everywhere
in the structure, as shown in Fig. 20, where temperature profiles
for selected models are presented. When the physical conditions
suitable for carbon burning ignition are achieved at the center,
the released nuclear energy is stored locally, due to the high
electron degeneracy; a convective core forms and grows
rapidly until convection reaches the surface. The thermonuclear runaway
will ultimately proceed on a dynamical time scale, causing the
disruption of the white dwarf as a type I supernova.

\placefigure{fig20}

In the second case ($\tau=10^3$ yr), the GWR-induced
compression also produces a homologous increase in the temperature
level throughout the structure. But, superimposed on this increase
is an additional increase in layers near the surface, where a local
temperature maximum forms (see Fig. 21). This secondary maximum occurs in part
because
matter near the surface, being less electron degenerate, has a larger
specific heat. Additionally, however, the evolution of the model
occurs on a time scale smaller than the thermal diffusion time scale 
in surface layers, so the thermal energy released into layers near
the surface remains localized in these layers. 
In any case, the evolution of the model is totally dominated by the
homologous compression due to GWR-induced braking, and, just as in the
first case, the thermonuclear runaway which has begun at the center
will evolve into an explosion of supernova proportions.

\placefigure{fig21}

Thus, the final outcome in both cases is a SN Ia event, independent of the
assumed GWR-braking efficiency.
The only effect of a variation in the braking efficiency over a substantial
range is a modest variation in the physical characteristics of matter when
carbon burning first develops into a thermonuclear runaway at the center. 
This is made clear in Fig. 22, where we show the evolution of central
characteristics in the $\rho - T$ plane for the two discussed models.

\placefigure{fig22}

\section{Final Remarks}

In this work, we have explored the lifting effect of rotation on the evolution
of a CO WD accreting CO-rich matter from the thick disk that is initially
formed during the merger of two close binary CO WDs. We find that rotation
induces 
a cooling of the accreting structure to the extent that the final outcome
of the accretion process can be significantly modified relative to the case
when rotation is neglected. For a fixed value of \mdot and an
initial angular velocity of the WD larger than a critical value,
the accreted matter produces a large enough thermal energy excess in outer 
layers that the star itself expands to a larger radius. The induced
expansion leads to a reduction of the surface gravity to the extent that the
accreting structure becomes gravitationally unbound (Roche instability)
and no more matter can be added. We have shown that this occurrence depends
on both the accretion rate and the initial angular velocity. For large values
of both \mdot and \o0,
the accreting WD becomes gravitationally unbound when only a small amount
of mass has been added ($\Delta M_{\rm acc}\le 0.1$ \msunend ), whereas, for
small
values of both \mdot and \o0, the WD attains a total mass close to 1.4 \msun
before the Roche instability sets in. We have also shown that behavior
of this sort depends critically on the angular momentum carried by the
accreted matter. For a fixed accretion rate, the larger the specific angular
momentum of the accreted matter, the earlier does the Roche instability occur
(the WD spins up more rapidly). 

The behavior we describe is a consequence of the fact that the thermal
diffusion time scale is very sensitive to changes in the assumed
rotation rate (increasing with increasing rotation rate) whereas the
compressional heating time scale is almost independent of such changes;
the rate of transfer toward inner regions of the energy excess produced
in outer regions by accreted matter becomes smaller and smaller with respect
to the standard case (without rotation) as the rotation rate is increased.
As a consequence, for a given accretion rate and WD mass, the density and
temperature at the center are smaller, the larger the initial angular velocity.
In addition, we find that, if the specific angular momentum of the accreted
matter is equal to that in the outer layers of the WD, 
all the computed models become gravitationally unbound well before the 
physical conditions suitable for carbon burning ignition are attained anywhere
in the WD structure. 

For small values of both initial angular velocity and accretion rate
(namely, $\omega_0 < 0.3 \omega_{\rm crit,0} =\sim 0.18$ rad/s and
\mdot $ \le 10^{-7}$\msun\yrm1 ),
the accreting model becomes a very fast rotator ($\omega_{\rm f}\sim 1$ rad/s)
when the total mass is of the order of 1.4 \msun and $\gamma$ 
($=E_{\rm rot}/E_{\rm g}$) becomes of
the order of the critical value ($\gamma_{\rm cr}\sim 0.1$). Then, the
accreting WD adopts an elliptical shape and begins to emit gravitational waves
which carry away both angular momentum and rotational energy. The important
point is that the braking induced by GWR produces a contraction of the whole
structure and, as a consequence, the WD spins up in such a way that the
$\gamma$ parameter continues to be of the order of the critical value. 

In this way, the evolution of the structure is totally driven by the homologous
compression induced by the GWR-braking process. According to our results,
independently of the efficiency of braking, which means independently of the
time scale on which braking occurs, the physical conditions suitable for an
explosive carbon-burning ignition are attained at the center of the WD,
so that a SN Ia event occurs. 

For models with large \o0 and \mdot, the Roche
instability occurs when the total mass of the WD is smaller than 1.4 \msun ,
the
exact value depending on the initial values of both the accretion rate and the 
angular velocity. But this result has been obtained by assuming that the
accretion rate is strictly constant; in the real world, the WD will continue
to accrete, but at a smaller rate. As the Roche instability is approached,
the accretion rate drops, the angular momentum of the WD increases
less rapidly, and the rate of producing an energy excess in the surface layers
decreases. Thus, a self-regulatory mechanism prevents the gravitational
instability from actually taking place. Rotation acts as a tuning mechanism
for the accretion rate in such a way that the accreting structure can
continue to accept mass even in the case of a very large initial accretion
rate. We address the details of this topic in a forthcoming paper. For now,
it is sufficing to observe that, although the accretion rate decreases
due to the lifting effect of rotation, the total angular momentum and,
as a consequence, the angular velocity of the WD must increase; the WD will
become a fast rotator. Gravitational wave radiation becomes competitive 
with thermal diffusion when the angular velocity of the WD attains
a value of the order of 1 rad s$^{-1}$, and, thereafter, GWR-induced
braking leads to the physical conditions suitable for carbon-burning
ignition in the center and, hence, to a SN Ia event.

\section{Acknowledgements}
We thank the referee for pointing to appropriate work in the literature
and for persuading us to examine various time scales more closely.





\begin{deluxetable}{ccccc}
\tablecaption{Results of experiments with standard conditions on
a 0.8\msun CO WD accreting at different accretion rates. $\dot{M}$ is the
accretion rate (in \msun\yrm1), $M_{\rm f}$ (in \msunend) is the total mass
at carbon ignition, $M_{\rm ig}$ is the mass coordinate where carbon burning
is ignited (in \msunend), $T_{\rm ig}$ (in $K$) and $\rho_{\rm ig}$
(in ${\rm g\ cm}^{-3}$) are, respectively, the temperature and the
density at the ignition point.
\label{table1}
}
\tablehead{
\colhead{$\dot{M}$}         &
\colhead{$M_{\rm f}$}           &
\colhead{$M_{\rm ig}$}          &
\colhead{$\log(T_{\rm ig})$}    &
\colhead{$\log(\rho_{\rm ig})$} 
}
\startdata
$        10^{-5}$ & 1.037 & 0.899     & 8.785 & 5.977 \nl
$5\times 10^{-6}$ & 1.193 & 1.134     & 8.773 & 6.242 \nl
$3\times 10^{-6}$ & 1.315 & 1.297     & 8.772 & 6.390 \nl
$2\times 10^{-6}$ & 1.383 & 0.000\tablenotemark{a)} & 8.295 & 9.565 \nl
$        10^{-6}$ & 1.380 & 0.098     & 8.381 & 9.371 \nl
$5\times 10^{-7}$ & 1.375 & 0.000     & 8.348 & 9.392 \nl
$        10^{-7}$ & 1.380 & 0.000     & 8.287 & 9.505 \nl
$5\times 10^{-8}$ & 1.383 & 0.000     & 8.237 & 9.576 \nl
$        10^{-8}$ & 1.395 & 0.000     & 8.101 & 9.705 \nl
\enddata
\nl
\tablenotetext{a)} {During the subsequent evolution of this model, the mass
coordinate where the nuclear energy production is maximum jumps to the
point $M=1.375$ \msun just below the surface.}
\end{deluxetable}

\newpage
\figcaption[fig1]{Chemical profiles of ${^{12}C}$ and  ${^{16}O}$
in a 0.8 \msun model at the beginning of the cooling sequence as obtained by
evolving a pure He star ({\em panel a}) and as it becomes after the 
re-homogenization made according to Salaris et al. (1997) ({\em panel b}) 
(see text).\label{fig1}}
\figcaption[fig2.eps] {Time evolution of the thermal diffusion time scale
(solid lines) and of the compressional heating time scale (dashed lines) 
for all of the computed models as listed in Table 1. \label{fig2}}
\figcaption[fig3.eps]{Time evolution of the central temperature (solid line),
the thermal diffusion time scale (dashed line), the mass coordinate where the 
maximum temperature is located (long dashed line) for the model with 
$\dot{M}=10^{-8}$ \msun\yrm1 (see text). All the quantities have been 
normalized to their maximum value. \label{fig3}}
\figcaption[fig4.eps]{Evolution of the temperature and of ratios of several
time scales in the interior of a model accreting at the rate $10^{-5}$
\msun \yrm1. Here, $\tau_{\rm diff,loc}$ is a local diffusion time scale,
$\tau_{\rm comp}$ is a local compressional time scale, and $\tau_{\rm evo}$ is
the time elapsed from the beginning of the accretion experiment. \label{fig4}}
\figcaption[fig5.eps]{Same as Fig. 4, but for $\dot{M}=10^{-8}$ \msun \yrm1.
 \label{fig5}}
\figcaption[fig6.eps]{Evolution in the HR diagram of all of the computed models
listed in Table 1 (see text). Attached to each line is the corresponding value 
of $\dot{M}$ (in \msun\yrm1 ). \label{fig5}}
\figcaption[fig7.eps]{Time evolution of the surface radius for all of the
models listed in Table 1 (see text). Attached to each line the corresponding
value of $\dot{M}$ (in \msun\yrm1 ). \label{fig7}}
\figcaption[fig8.eps]{Evolution in the $\rho - T$ plane of the center of the
0.8 \msun CO WD accreting matter at various $\dot{M}$, as labeled
inside the figure (in \msun\yrm1 ) (see Table 1). The long-dashed curve labeled
" $\varepsilon_{\nu} = \varepsilon_{cc}$ " is the ignition curve. 
\label{fig8}}
\figcaption[fig9.eps]{Temperature profiles for several selected structures 
during the accretion process for different values of the accretion rate, as
labeled inside each panel. \label{fig9}}
\figcaption[fig10.eps]{Density profiles for the same models as in Fig. 9. 
\label{fig10}}
\figcaption[fig11.eps]{Profiles in the $\rho-T$ plane of models which
have attained the physical conditions suitable for carbon ignition
as a consequence of mass accretion at different $\dot{M}$, as labeled
inside the figure (in units of \msun \yrm1).
\label{fig11}}
\figcaption[fig12.eps]{Initial value of the thermal diffusion time scale 
$\tau_{\rm diff}$ (in yr) as a function of the initial angular velocity (in 
units of $\omega_{\rm crit,0}$). \label{fig12}}
\figcaption[fig13.eps]{Time evolution of the surface radius for models
rotating with different initial angular velocities as labeled inside the figure
(in units of $\omega_{\rm crit,0}$). Each panel refers to a different value
of the accretion rate as labeled. For comparison the evolution of the standard
model without rotation (dashed line) is also shown. \label{fig12}}
\figcaption[fig14.eps]{Evolution of the diffusion time scale in the outer zone 
({\em panel a}), the surface temperature ({\em panel b}),
and the total radius ({\em panel c}) of the model with 
$\dot{M}=10^{-8}$ \msun\yrm1 and $\omega_0=0.5\omega_{\rm crit,0}$. In
{\it panel d}, the evolution of the white dwarf angular velocity (lower
curve) is compared with that of the critical angular velocity at the surface
(upper curve).
\label{fig14}}
\figcaption[fig15.eps]{The region in the parameter space $\dot{M} - \omega_0$ 
(shaded zone) for which the rotating models accreting CO-rich matter 
become gravitationally unbound when the total mass of the WD approaches 1.4
\msun (see text). \label{fig15}}
\figcaption[fig16.eps]{The same as in Fig. 14, but for the case 
$\dot{M}=10^{-8}$ \msun\yrm1 and $\omega_0=0.1\omega_{\rm crit,0}$.
\label{fig16}}
\figcaption[fig17.eps]{Temperature profiles of selected structures 
during the evolution of the rotating model with $\omega_0=0.1\omega_{\rm
crit,0}$ 
and accreting matter at $\dot{M}=10^{-8}$ \msun\yrm1. The labels attached to
each line provide the time elapsed from the beginning of the accretion process.
\label{fig17}}
\figcaption[fig18.eps]{Evolution of the surface radius for models accreting
CO-rich matter at $\dot{M}=10^{-7}$\msun \yrm1 with different values of the
initial angular velocity computed by assuming that the accreted matter 
does (dashed lines) or does not deposit (solid lines) angular momentum 
(see text).  \label{fig18}}
\figcaption[fig19.eps]{The same as in Fig. 15, but for rotating models which
accrete CO-rich matter at rest (see text). \label{fig19}}
\figcaption[fig20.eps]{Temperature profiles for some selected structures 
during the braking stage of a rotating $\sim 1.42$ \msun CO WD. The braking
time scale has been set equal to $10^7$ yr (see text).\label{fig20}}
\figcaption[fig21.eps]{Temperature profiles for some selected structures 
during the braking stage of a rotating $\sim 1.42$ \msun CO WD. The braking
time scale has been set equal to $10^3$ yr (see text).\label{fig21}}
\figcaption[fig22.eps]{Evolution in the $\rho - T$ plane of the center of
models accreting CO-rich matter at $\dot{M}=10^{-8}$ \msun\yrm1. The
dashed line describes the evolution of the standard model ($\omega=0$),
whereas the solid lines describe the evolution of two rotating models with
$\omega_0=0.1\omega_{\rm crit,0}$ and two different GWR-braking times,
one with $\tau=10^3$ yr, the other with $\tau=10^7$ yr (see
text).\label{fig22}}
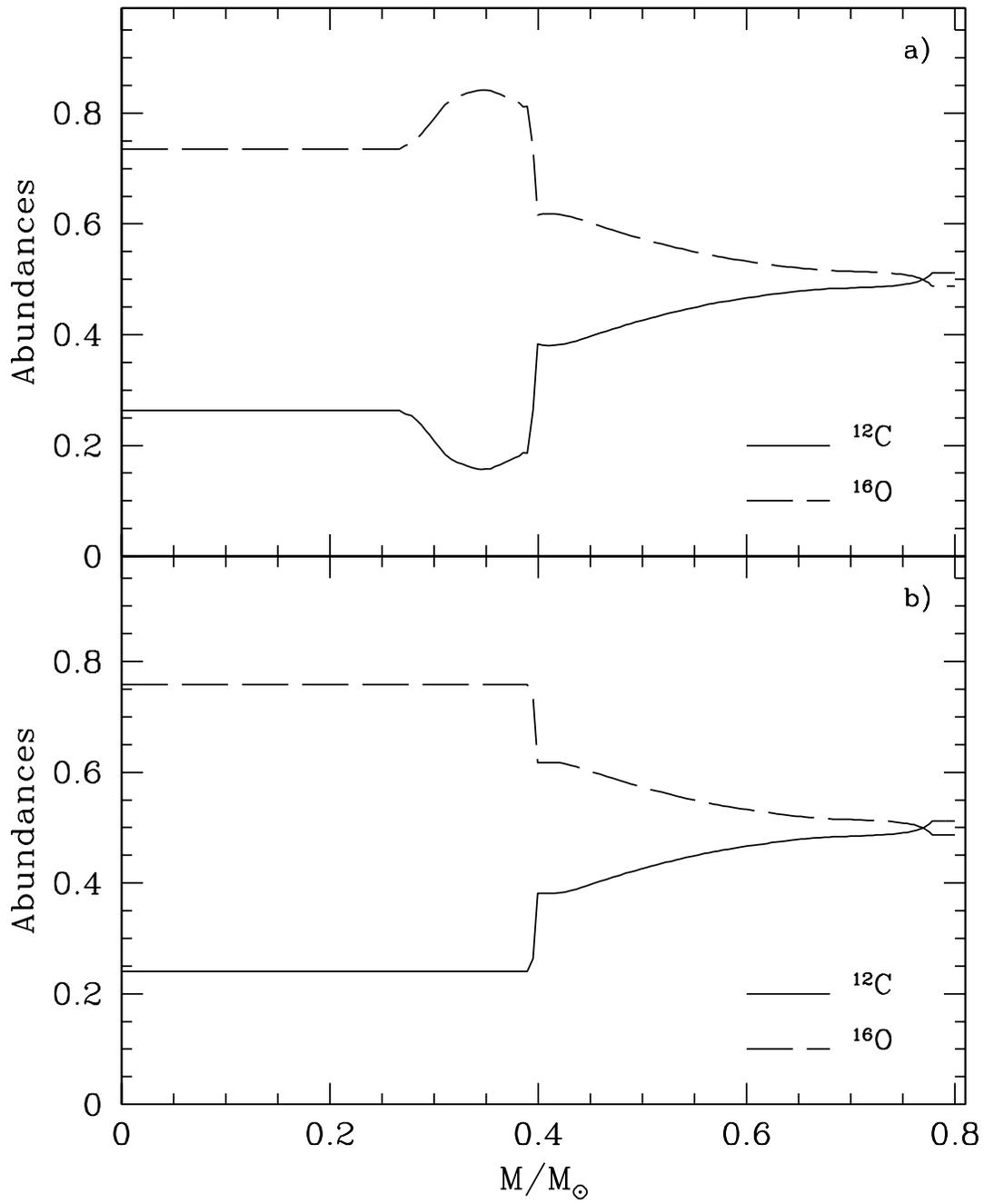
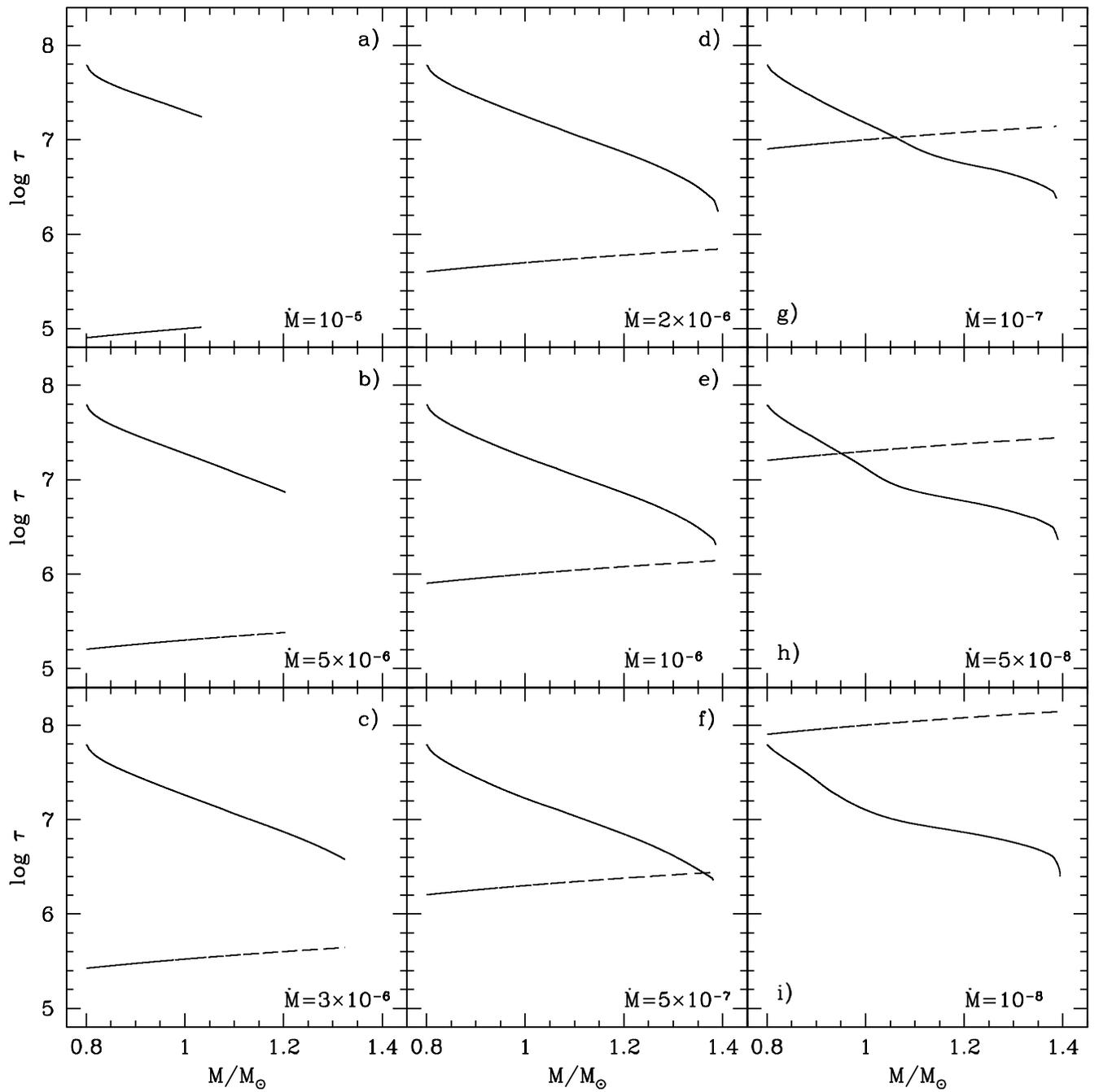
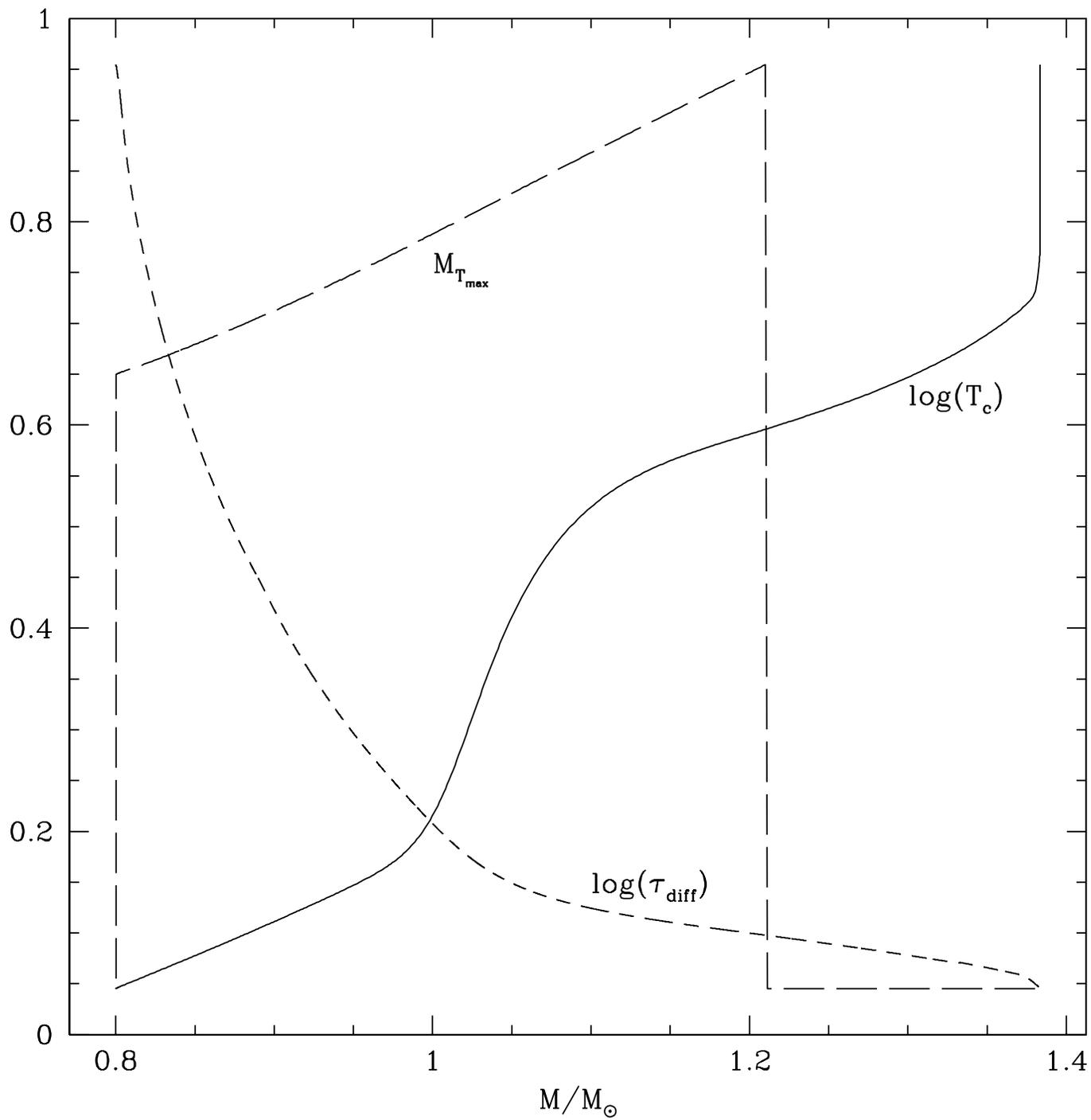
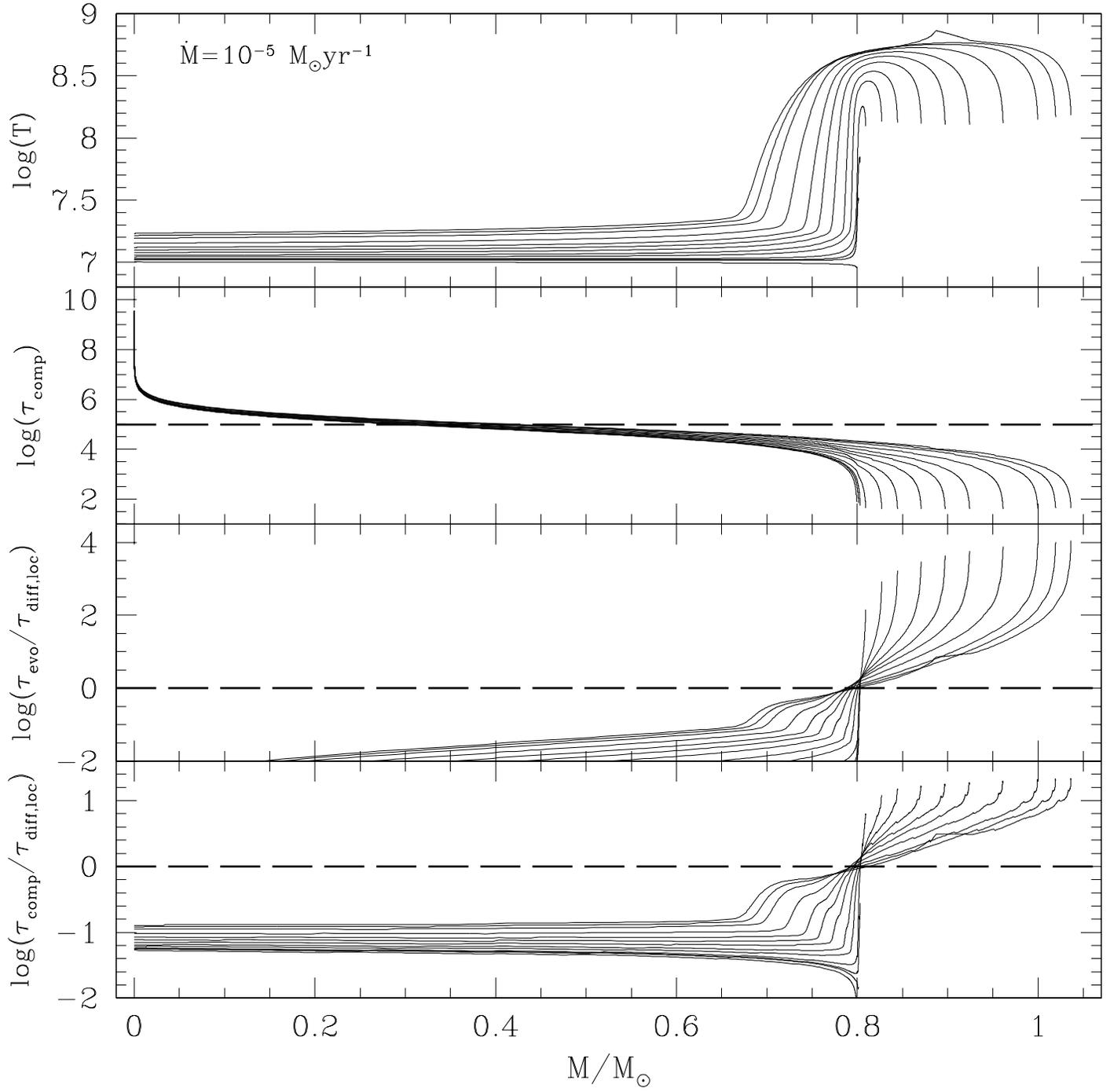
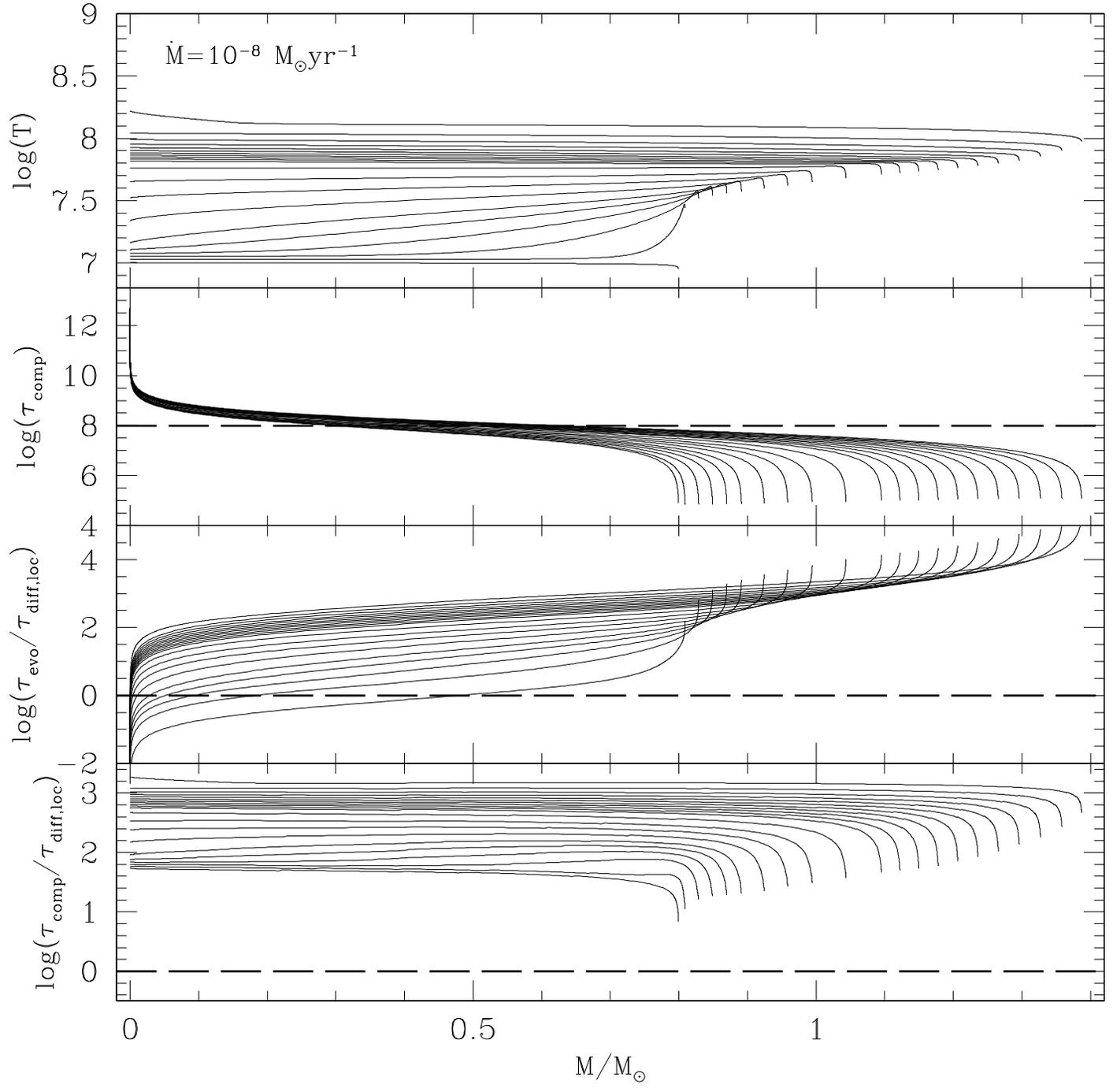
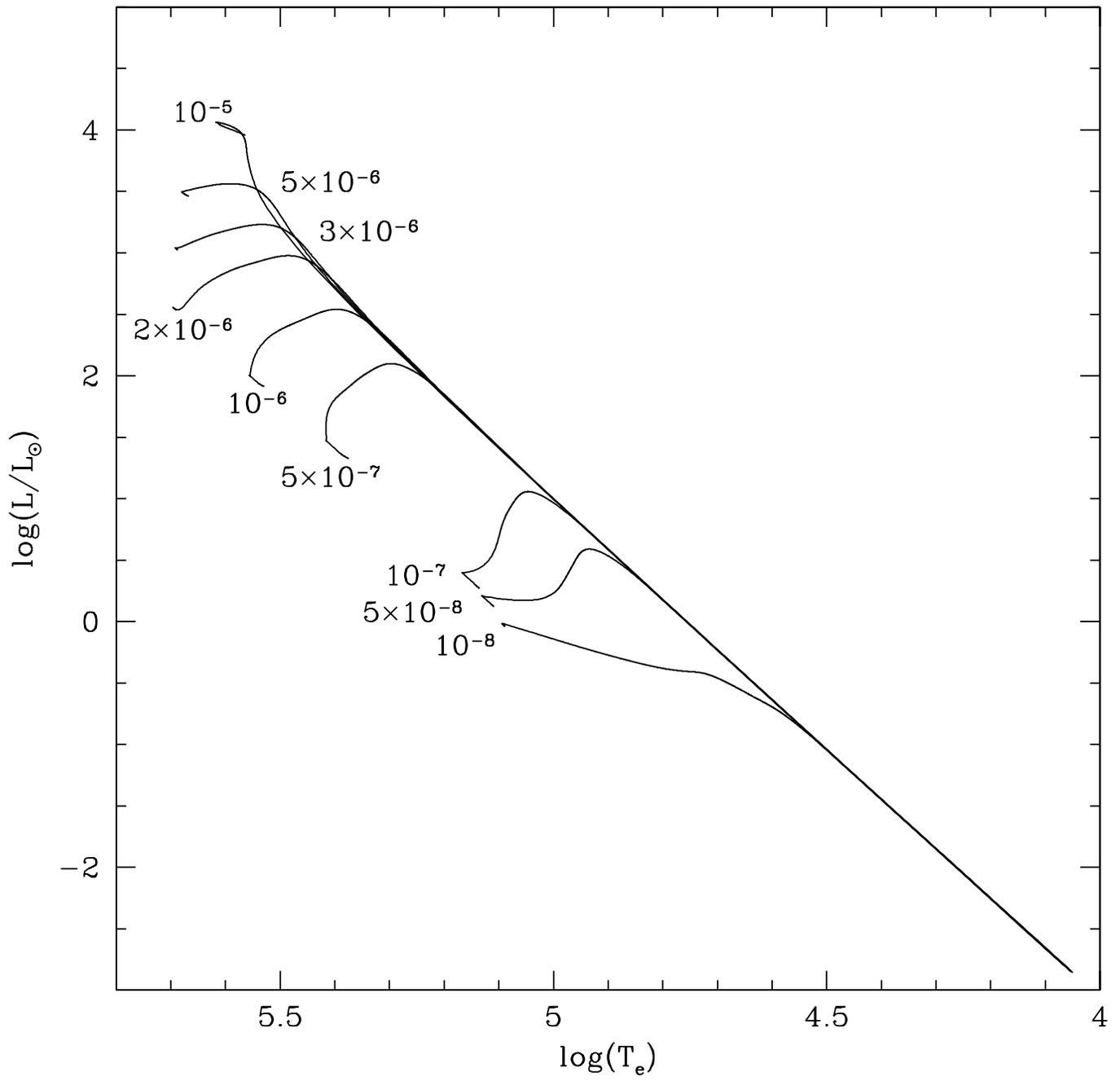
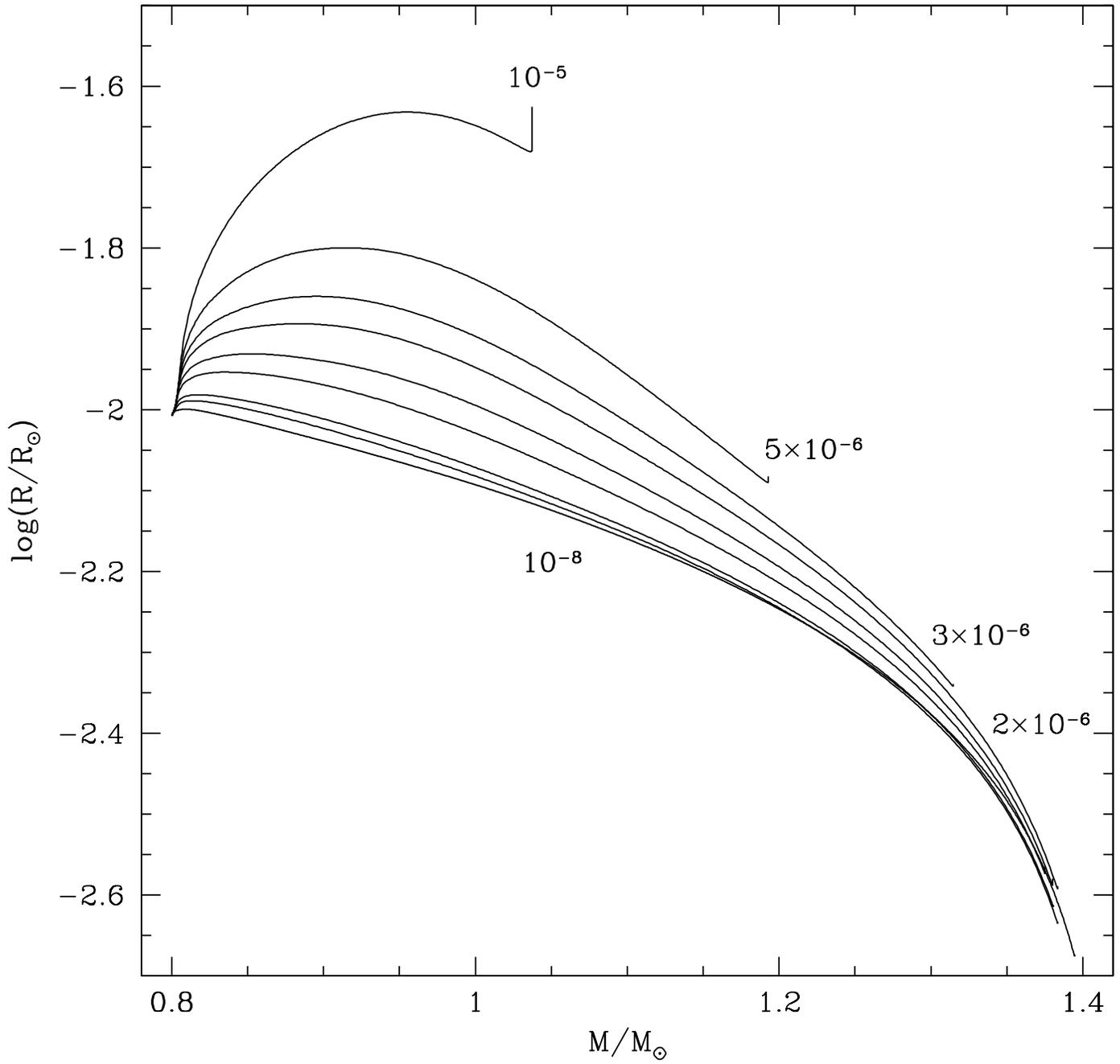
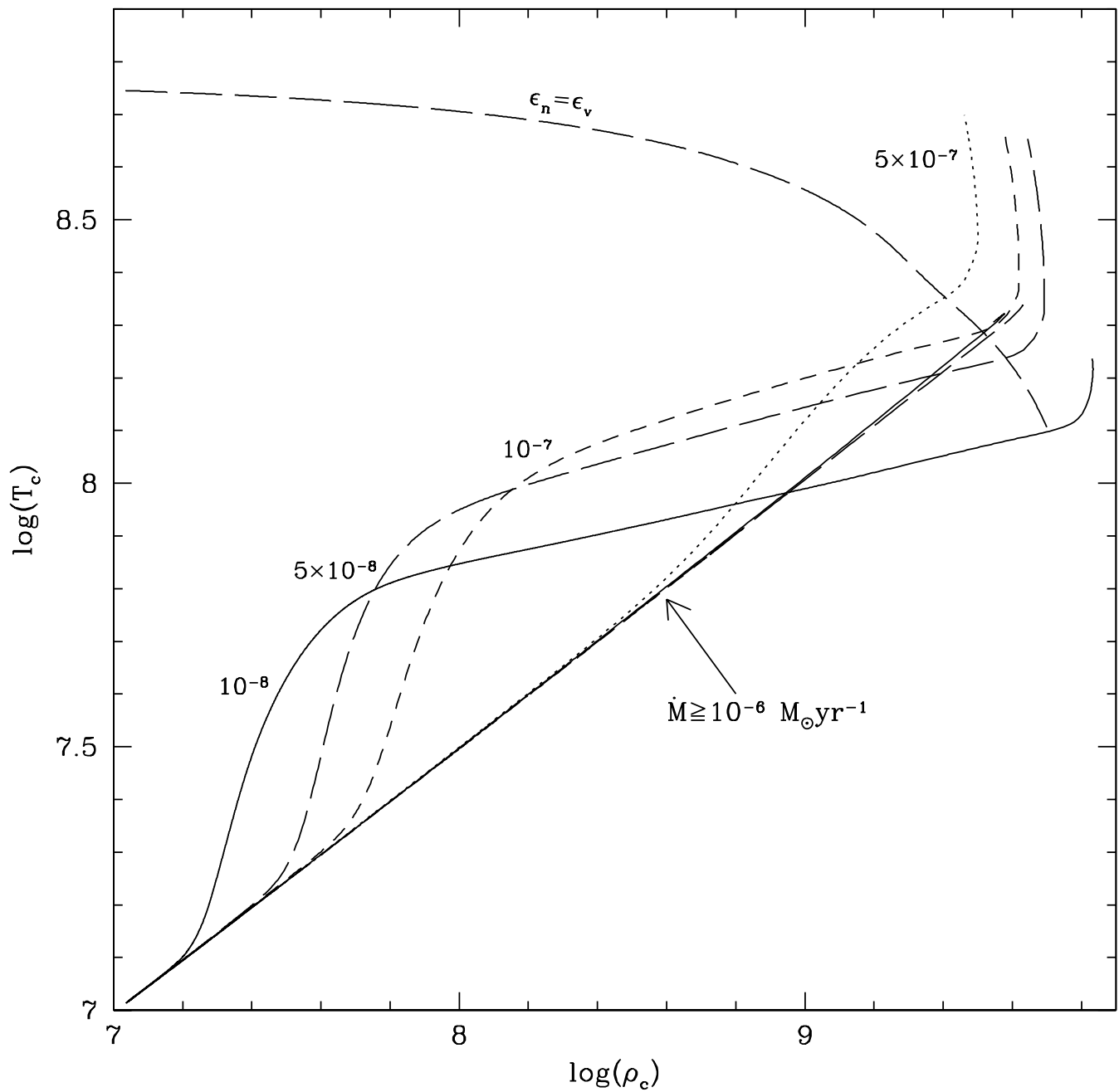
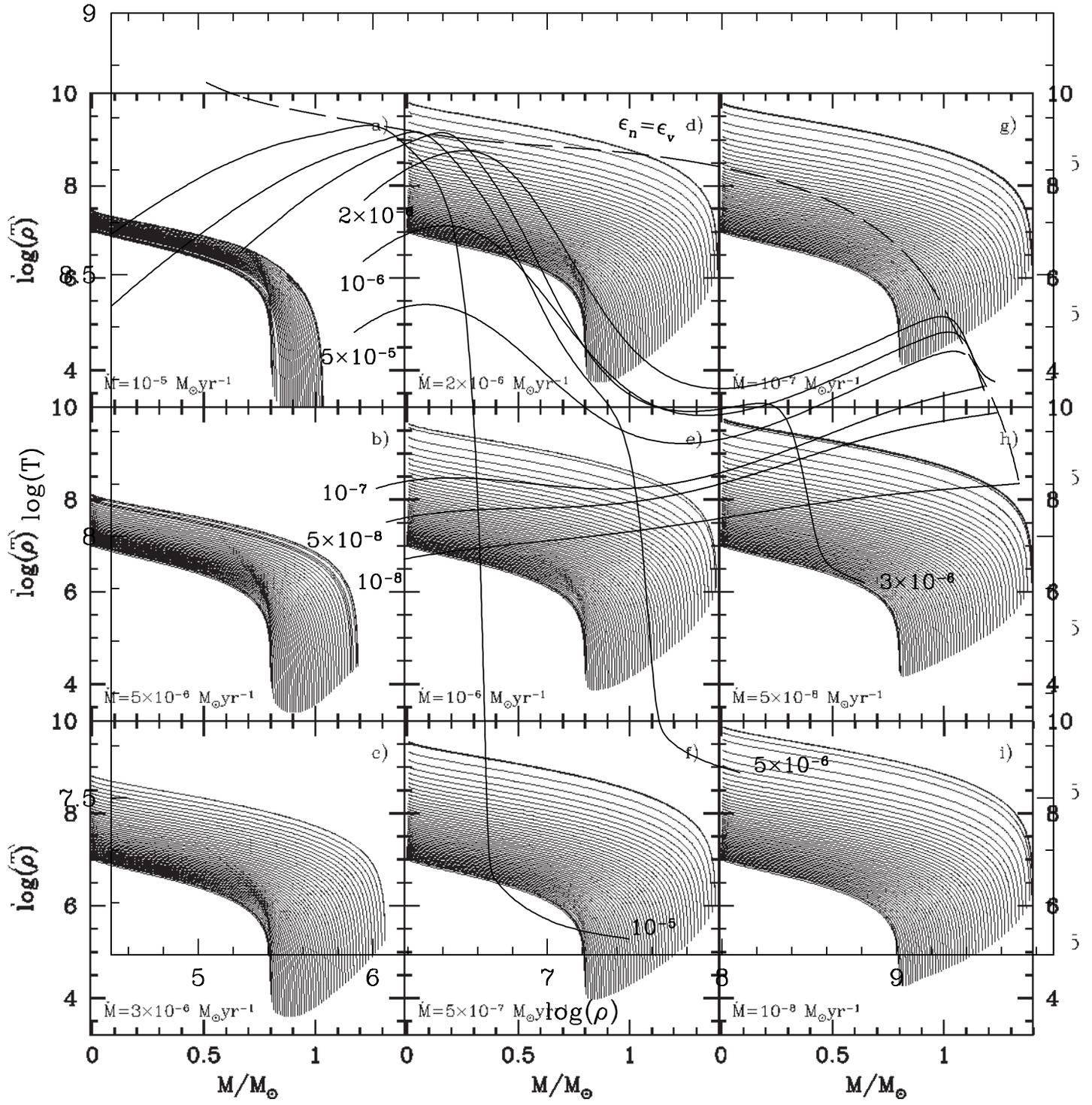
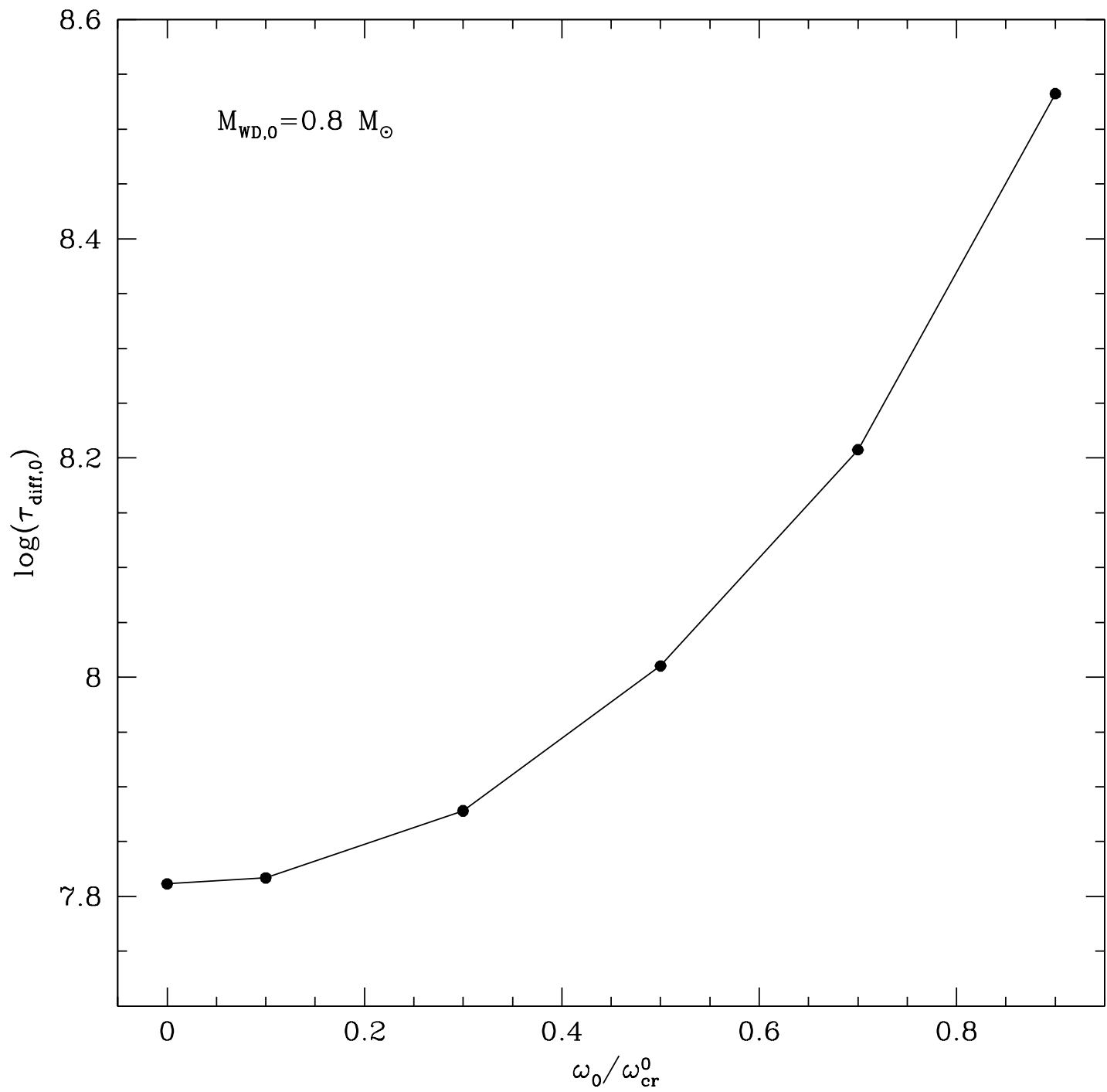
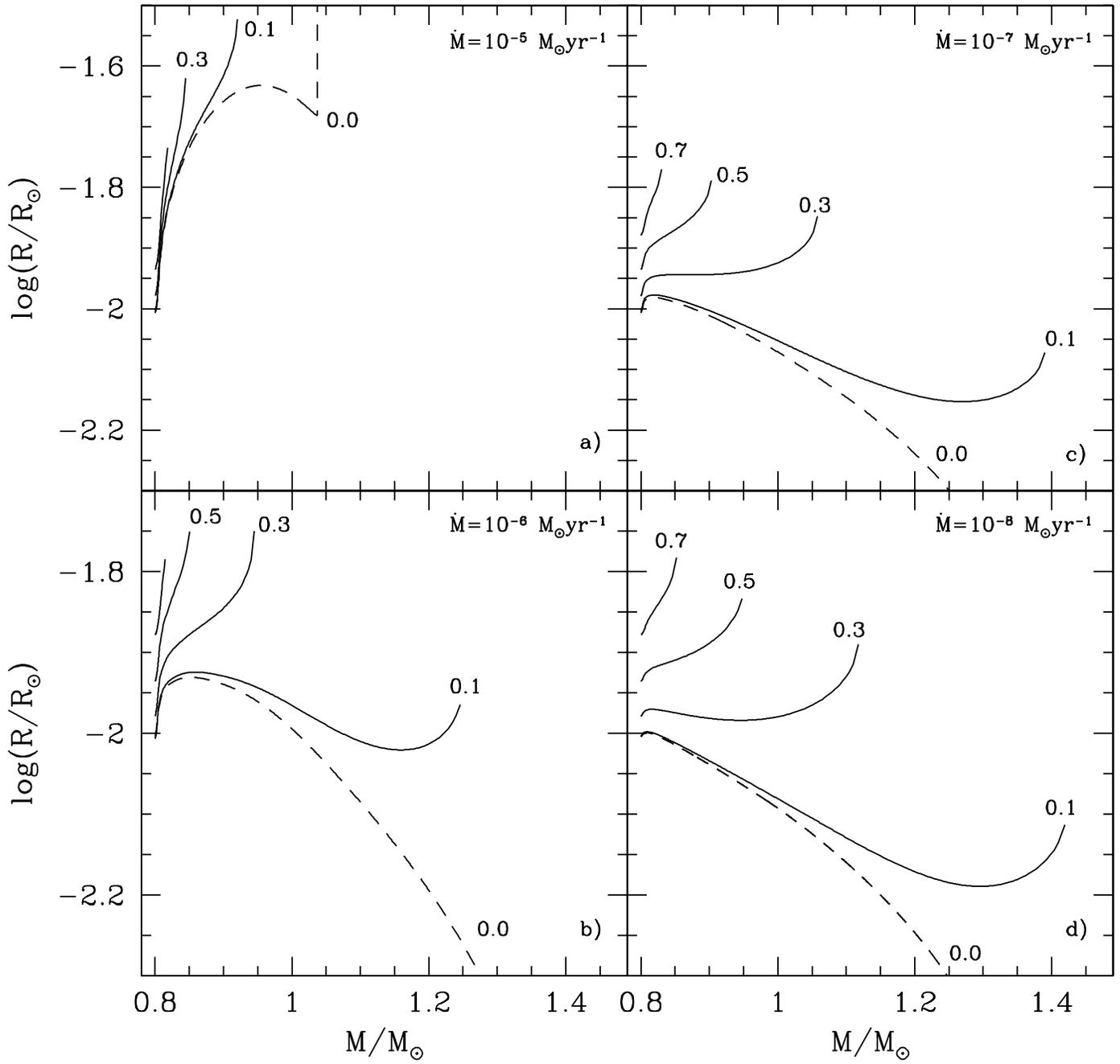
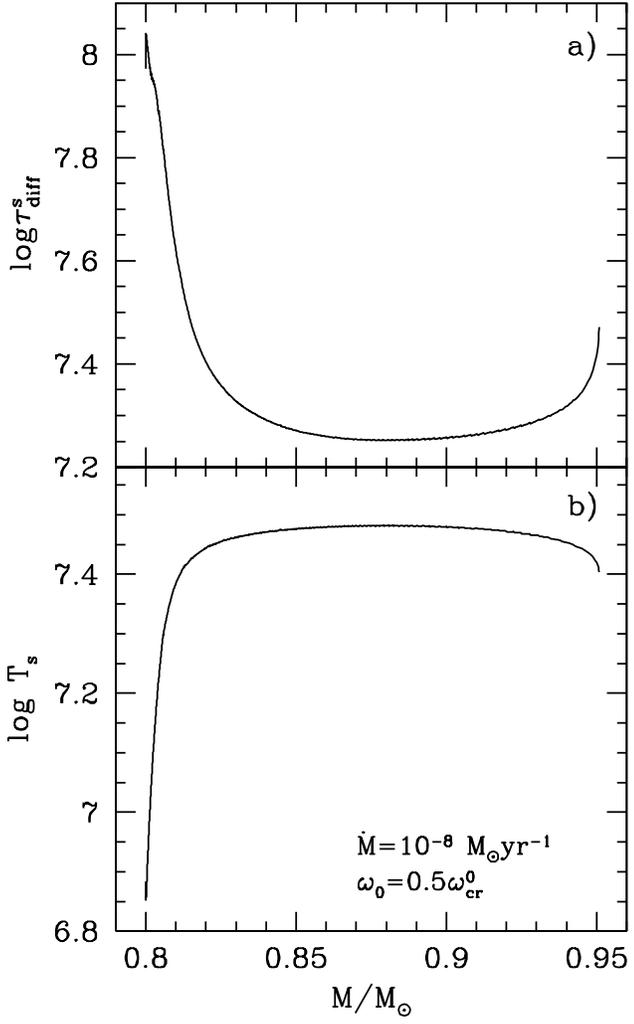
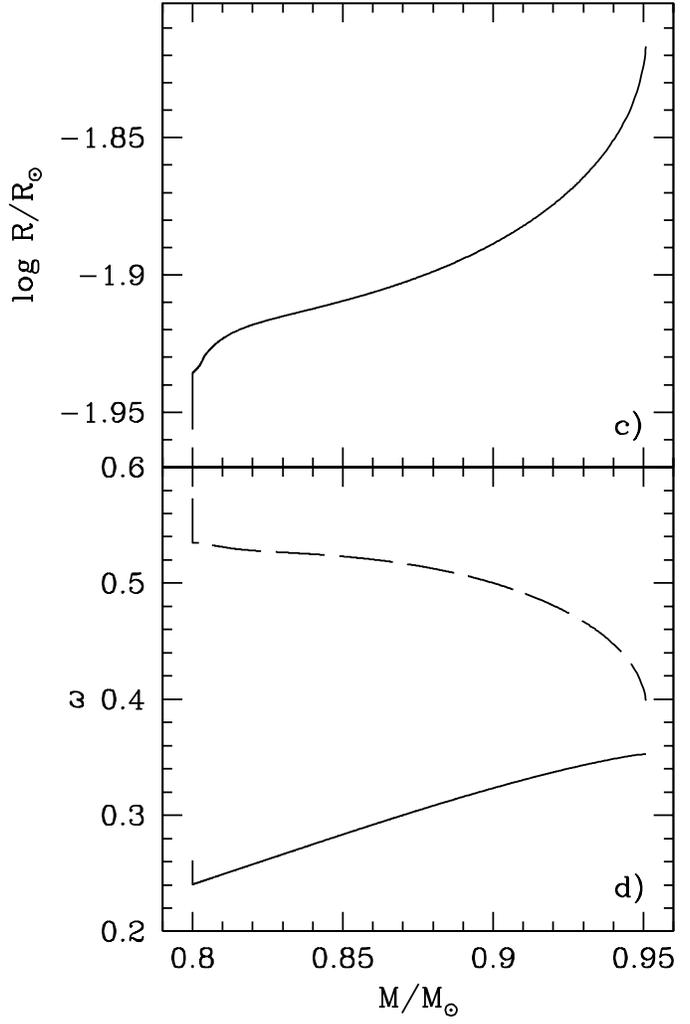
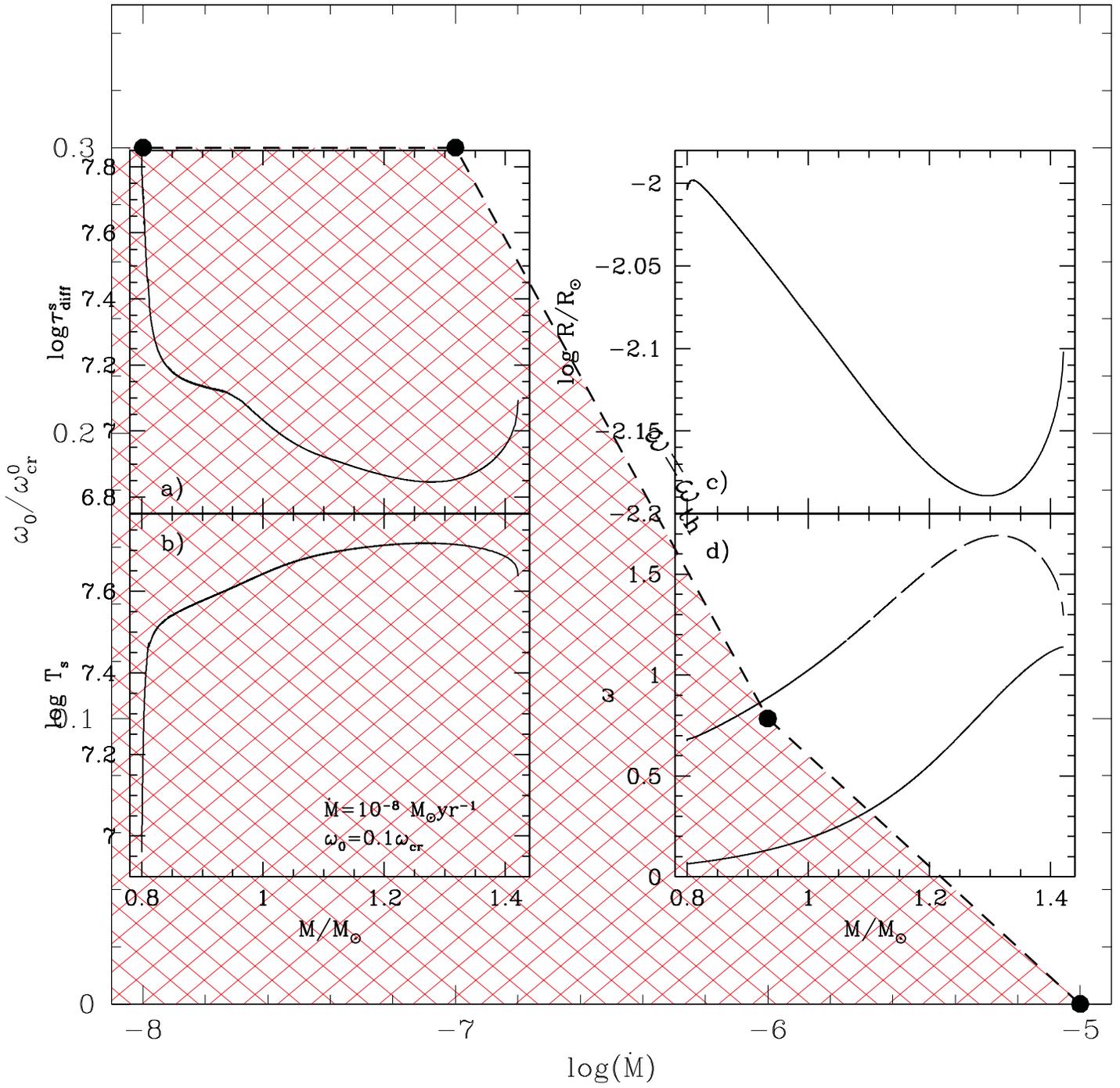
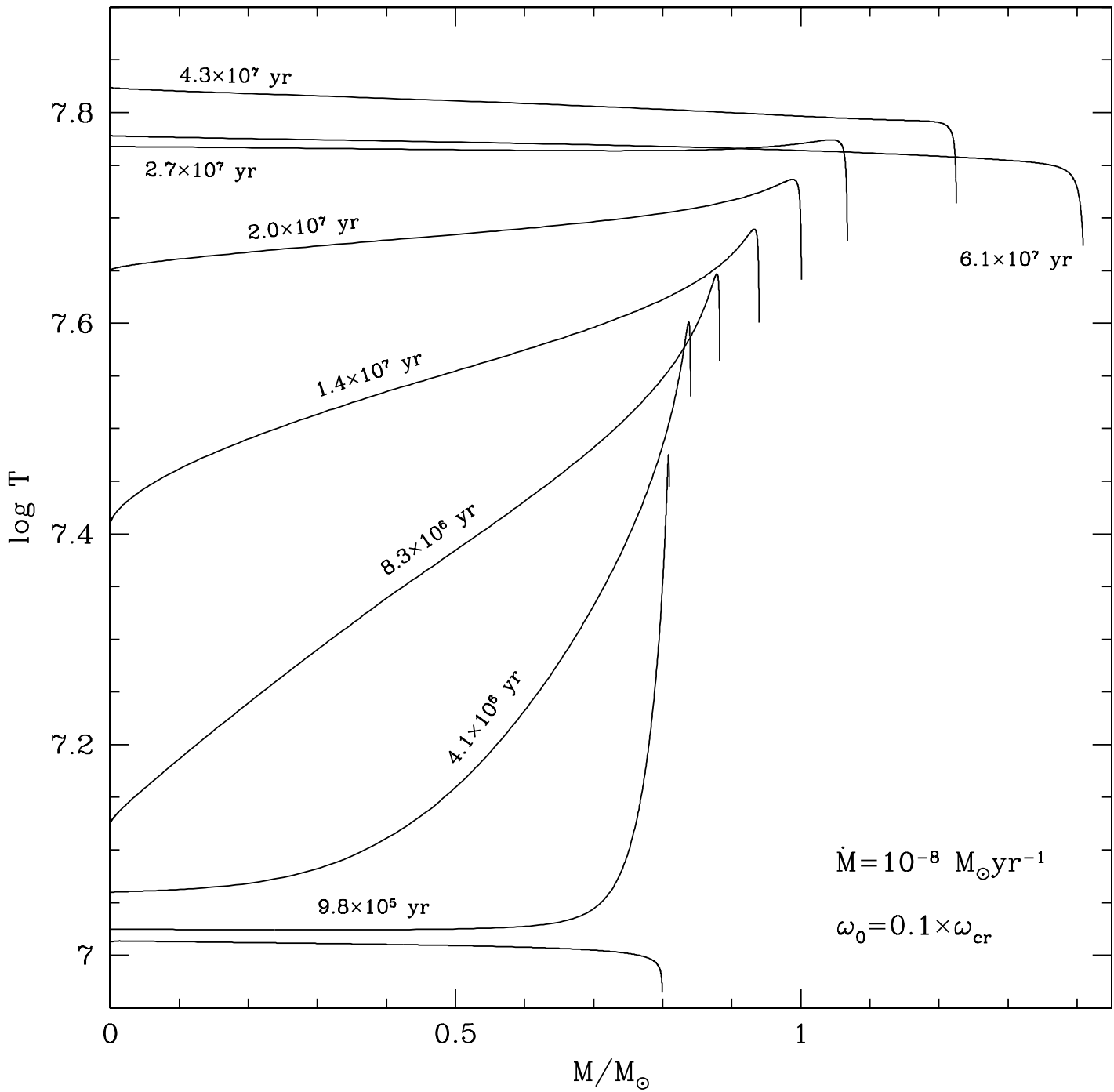
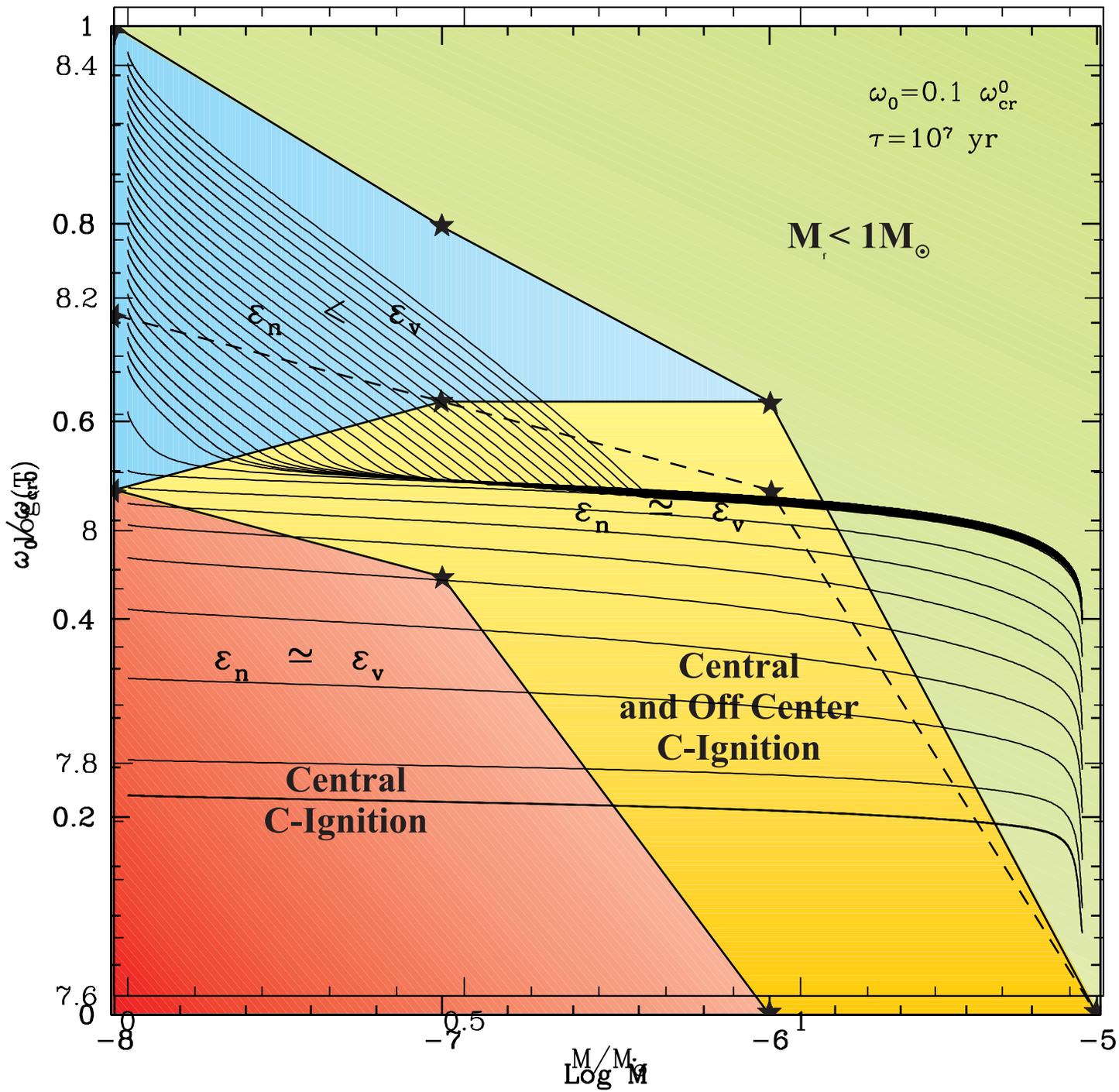
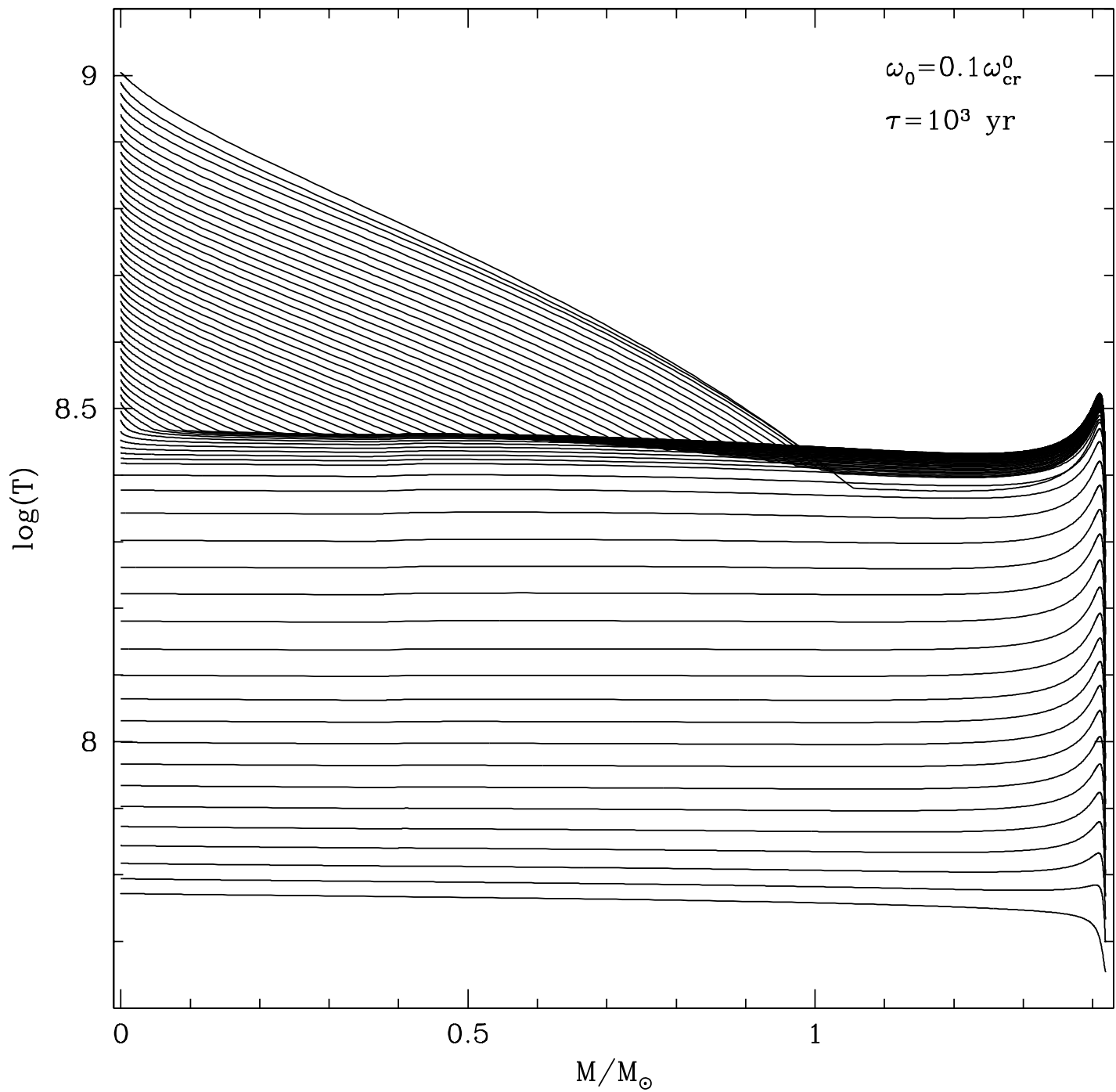
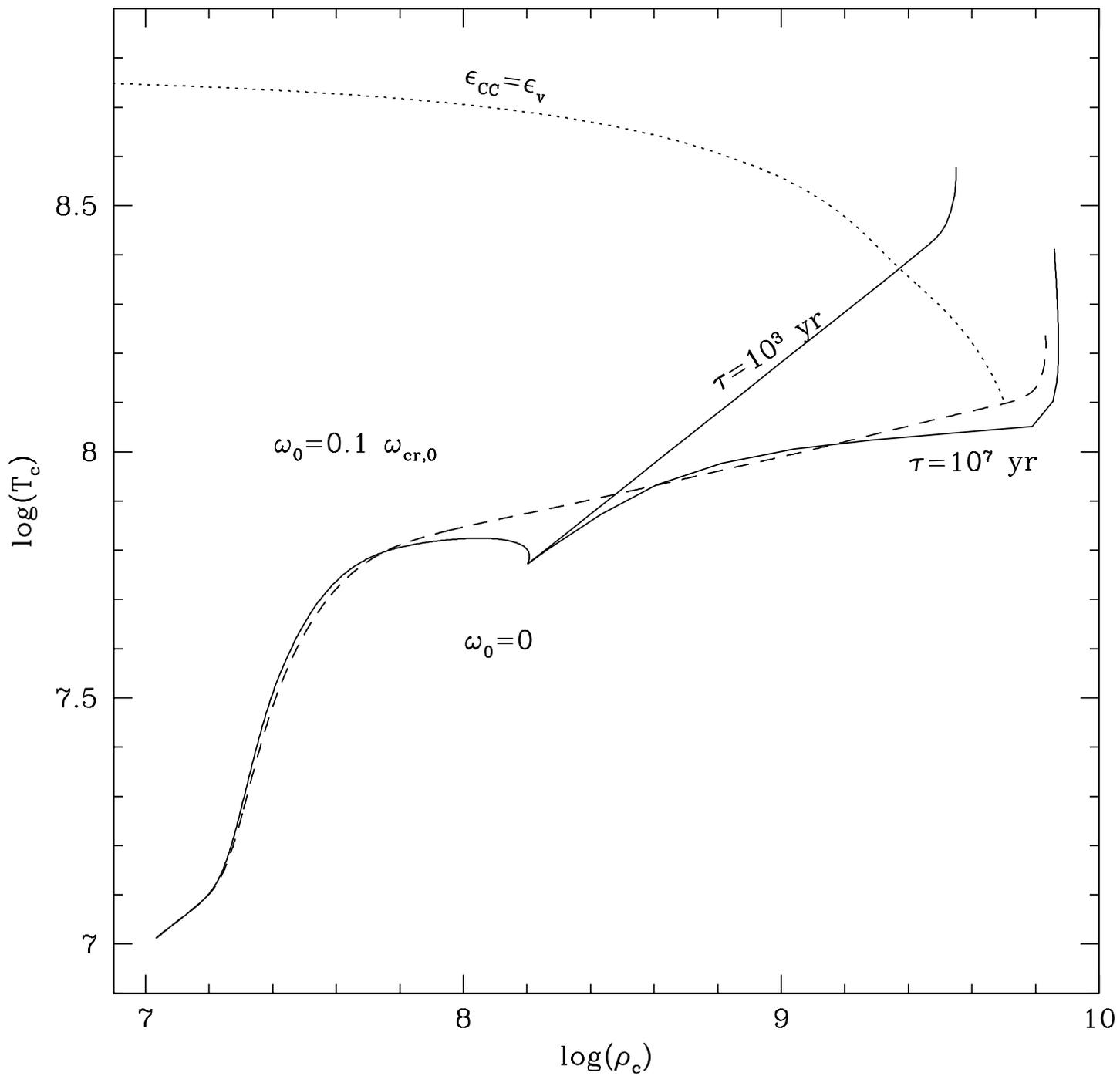
\end{document}